\documentclass[aps,amsmath,amsfonts,twocolumn]{revtex4-2}%
\usepackage{mathtools}
\usepackage{verbatim}
\usepackage{xr-hyper}
\usepackage{hyperref}
\usepackage{cleveref}
\usepackage{epsfig,color,amsfonts,amsmath,amsthm,comment,natbib}
\usepackage{color}
\usepackage{graphicx}
\usepackage{braket}
\usepackage{amsthm}
\usepackage{amssymb}
\usepackage{amsmath}
\usepackage{ragged2e}
\usepackage{ulem}
\usepackage{tikz}
\usepackage[caption=false]{subfig}

\DeclareMathOperator{\Tr}{Tr}

\newcommand{\ee}{{\rm e}}
\newcommand{\ii}{{\rm i}}
\newcommand{\Var}{\text{Var}}

\begin{document}

\title{Absorbing State Phase Transitions Beyond Directed Percolation in Dissipative Quantum State Preparation}

\author{Matthew Wampler}
\author{Nigel R. Cooper}
\affiliation{T.C.M. Group, Cavendish Laboratory, University of Cambridge, J.J. Thomson Avenue, Cambridge CB3 0US, United Kingdom \looseness=-1}

\begin{abstract}
We show that absorbing state phase transitions where the absorbing state itself exhibits long-range phase coherence can lead to critical behavior distinct from Directed Percolation.  To do this, we investigate a simple, purely dissipative quantum reaction-diffusion model, which may also be viewed as a dissipative quantum state preparation procedure for the (generalized) W-state with errors.  The `error' Lindblad jump operators preserve the W-state as a dark state, but nonetheless act to decohere the system and induce the phase transition. We find cases where the preparation protocol is either fragile or robust against weak error quantum jump rates and show that local remnants of the coherence persist in the decohering phase.  The distinct critical behavior stems from the spreading of coherence throughout the system at the critical point.  
\end{abstract}


\maketitle

\textit{Introduction}---The dynamics of non-equilibrium quantum many-body systems exhibit striking phenomena which would be rare, if not impossible, in classical or equilibrium settings.  Interest in this area has been especially spurred by recent experimental advances -- across a variety of platforms -- in the control of quantum systems at the atom-by-atom (or qubit-by-qubit) level \cite{Barreiro2011OpenTrappedIons,MULLER2012DissipativeBEC,Langen2015UltracoldNonEquilibrium,Kjaergaard2020SuperconductingQubits,Mivehvar2021CavityQED}.  

Often, evolution in these systems has been viewed as a competition between the coherent Hamiltonian dynamics which may contribute to the quantum nature of a system, and the decohering effects of coupling to the environment or measurements which destroy entanglement and other quantum signatures.  However, even in systems with no Hamiltonian dynamics, it is possible to generate entanglement solely through dissipation \cite{Verstraete2009DQC,Ippoliti2021MeasurementOnly}.  Additionally, the irreversible nature of dissipation allows for the one-way cooling of quantum systems to a desired steady state as well as provides the possibility for interactive control of a system through feedback.  These properties -- which allow dissipative dynamics to be especially robust to perturbation -- make it a key tool in quantum information \cite{Harrington2022EngineerDissipation}, showing up in quantum state preparation and stabilization \cite{Valenzuela2006StatePrep,Kraus2008DissipStatePrep,Ticozzi2012DStatePrep,Ticozzi2014DStatePrep,Didier2014StatePrep,Leghtas2015StatePrep,Liu2016StatePrep,Boutin2017StatePrep,nathan2024dissipativeGKP}, error correction \cite{Kapit2017errorcorrect,Reiter2017errorcorrect,Gertler2021errorcorrect,deNeeve2022errorcorrect,Sivak2023errorcorrect}, and as distinct models of universal quantum computation \cite{Verstraete2009DQC,Briegel2009MBQC}.

In condensed matter, the same experimental advancements in engineered coupling to the environment and quantum control have led to the discovery of new phases of matter and dynamical phase transitions which are unique to the non-equilibrium setting \cite{Walls1978DPT,DallaTorre2010DPT,Torre2013DPT,Klinder2015DPT,Fitzpatrick2017DPT,Skinner2019DPT}.  Non-equilibrium phase transitions are present in a wide variety of contexts \cite{Barabasi1995Surface,MAERIVOET2005Traffic,Harris1974Infect,Ravindranath2023ASPT,ODea2024ASPT,odea2024ASPTlongrange}.  
An important class of systems which exhibit such transitions are reaction-diffusion models, where particles (or reactants) move through a system by diffusion and react when they meet. 
 Here, the system may exhibit phase transitions to absorbing states (i.e. configurations which may be approached by the dynamics, but not left); such transitions typically belong to the Directed Percolation (DP) universality class \cite{HinrichsenReactionDiffusion}.  There has been a flurry of recent interest---on both theoretical \cite{sieberer2023universalitydrivenopenquantum,VanHorssen2015ReactDiff,Marcuzzi2016ReactDiff,Roscher2018RGcritical,Carollo2019ReactDiff,Gillman2019numericalcritical,Lin2021ReactDiff,Carollo2022quantumabsorbing,Perfetto2023ReactDiff,Perfetto2023ReactDiff2,Makki2024CliffordCriticalBehavior,Thompson2025QuantumASPT} and experimental \cite{Gutierrez2017ReactDiff,Helmrich2020QuantumRDExperiment,klocke2021QuantumRDExperiment,Chertkov2023QuantumDPExperiment} fronts---into quantum extensions to reaction-diffusion models, with continuing debate and many open questions surrounding precisely when and how quantum effects may induce novel critical behavior distinct from DP.   

In this paper, we show that critical behavior beyond DP can occur when the absorbing state itself has long range coherence.  This distinct behavior stems from dynamics associated with the spreading of quantum coherence throughout the system at the critical point.  To show these results, we provide a prototypical example of a quantum reaction-diffusion model with a transition to a long range coherent absorbing state.  We find, furthermore, this coherence may be induced entirely through dissipative means, with no Hamiltonian evolution.  The model may also be viewed as a dissipative quantum state preparation procedure---preparing the long range entangled absorbing state---with additional `error' environmental couplings that ruin the preparation protocol.  The absorbing state phase transition, from this perspective, may be viewed as a probe into the stability of purely dissipative quantum state preparation.

The model consists of local Lindbladian dynamics that includes the generalized W-state as a dark state.  These are a class of states given by 

\begin{gather}
    |W_{\text{Gen}}\rangle = \frac{1}{\sqrt{N}} \sum_{a=0}^{N-1} \ee^{\ii \phi_a} |\downarrow_0...\downarrow_{a-1} \uparrow_a \downarrow_{a+1} ... \downarrow_{N-1} \rangle
    \label{eq: General W}
\end{gather}
where $a$ labels the sites of the lattice, and various choices of phase $\phi_a$ give, for example, the W-state (all $\phi_a=0$) or vortex lattices (phase winding around closed loops).  
While the (generalized) W-state is a dark state of the Lindbladian (to be defined below), some terms are ``attractive'' in the sense that they drive the system towards the W-state (the W-state preparation protocol) while other terms are ``repulsive'' and drive states in the neighborhood of the W-state towards a different state (in this case, the maximally mixed state); see the end of the next section for a precise definition as well as supplementary material \cite{supp} for further details.  Competition between these terms induces a phase transition away from the absorbing W-state.  



\begin{figure}
    \centering
    \includegraphics[width=0.45\textwidth]{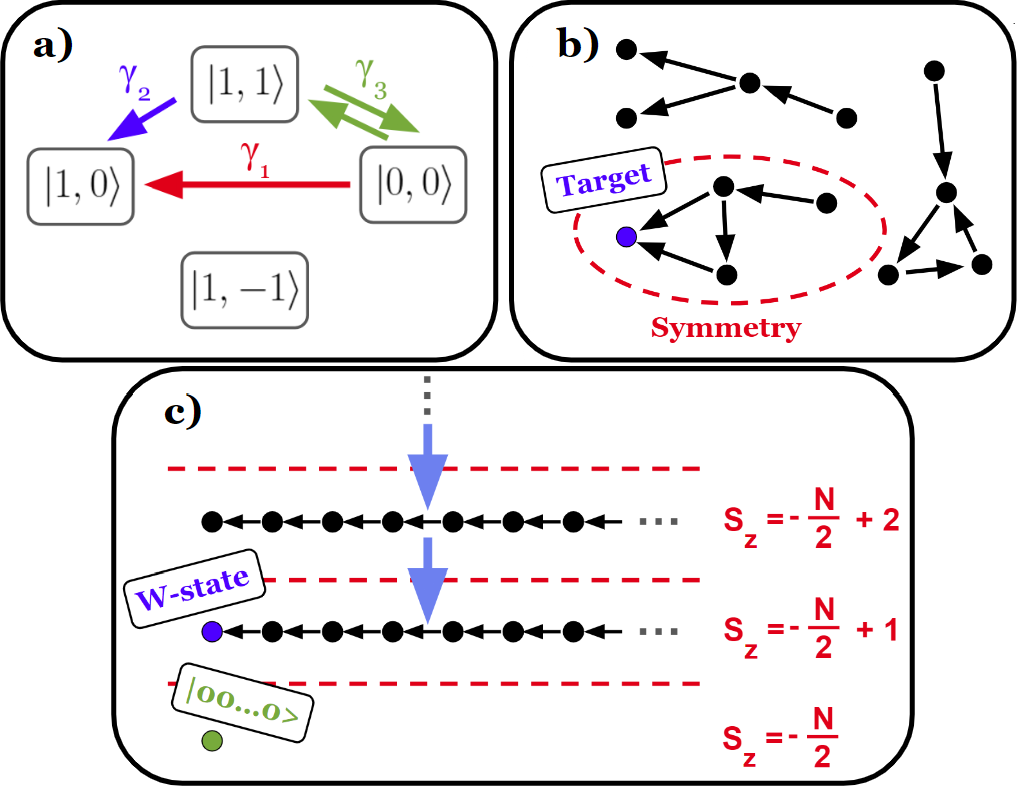}
    \caption{Our model and dissipative W-state preparation.  a) Quantum Jump operators for the model written in terms of total spin eigenstates $|s,m_s\rangle$ for 2-site pairs.  The $\gamma_1$, $\gamma_2$ terms prepare the W-state, while $\gamma_3$ terms induce a phase transition to a mixed state. b) Points represent states and arrows are quantum jumps in the Lindblad dynamics. c) Preparation of W-state using ladder between symmetry sectors.}
    \label{fig: DPT Model}
\end{figure}
\textit{The Model}---The model we consider is a lattice of two level systems where pairs of lattice sites are coupled (via a set of environments) to undergo dissipative evolution.  We restrict to the case well-described by Lindbladian dynamics, given by the master equation for the density operator $\rho$
\begin{gather}
    \Dot{\rho} = \sum_\ell \gamma_\ell \left( L_\ell \rho L_\ell^\dagger - \frac{1}{2} \left\{L_\ell^\dagger L_\ell,\rho \right\}\right).
    \label{eq: Lindbladian}
\end{gather}
The model is purely dissipative (with real $\gamma_\ell>0$), which is why no coherent, Hamiltonian evolution appears in \eqref{eq: Lindbladian}.

In particular, we consider 2-site jump operators grouped into the following four sets:
\begin{subequations}
\label{eq: All L}
\begin{gather}
    L_1 = \left\{|1,0\rangle_{a,b}\langle 0,0|_{a,b} \right\}, \label{eq: L1}\\
    L_2 = \left\{|1,0\rangle_{a,b} \langle 1,1|_{a,b} \right\}, \label{eq: L2}\\
    L_3 = \left\{|0,0\rangle_{a,b} \langle 1,1|_{a,b} \right\},  \label{eq: L3} \\
    L_{3'} = \left\{|1,1\rangle_{a,b} \langle 0,0|_{a,b} \right\}.  \label{eq: L3'} 
\end{gather}
\end{subequations}
We will consider two cases: all-to-all jump operators---with any pair of sites $a,b$---as well as nearest-neighbor (NN) where $a,b$ are on adjacent lattice sites (with coordination number $Z$).  We have written 2-site states in terms of total spin quantum numbers $|s,m_s\rangle$.

In the Lindbladian \eqref{eq: Lindbladian}, terms $L_1$, $L_2$, $L_3$, and $L_{3'}$ have coefficients $\gamma_1$, $\gamma_2$, $\gamma_3$ and $\gamma_{3'}$ respectively.  Throughout the work, we will take $\gamma_3 = \gamma_{3'}$.  The model is summarized in Fig. \ref{fig: DPT Model}a. 

To see that this Lindbladian represents a quantum reaction-diffusion model, view the spin-up excitations as hard-core bosonic particles.  From this viewpoint, $L_1$ conserves particle number (total $S_z$) and---since the dissipative evolution favors states where a particle is in a coherent superposition on neighboring sites---will tend to spread the particle throughout the system, i.e. it is the quantum analog of the `diffusion' term.  On the other hand, the rest of the jump operators correspond to prototypical reaction processes. 
 Namely, `coagulation' terms $L_2$ and $L_3$ correspond to processes where two particles collide (i.e. are on adjacent sites) and decay down to a single particle state, while `branching' terms $L_{3'}$ correspond to Lindblad jumps from a single particle state to a two particle state.  Two-site dissipative dynamics similar to that generated by the quantum jumps above have been realized across a variety of experimental platforms, for example \cite{Barreiro2011TwoSiteJump,Shankar2013TwoSiteJump,Leghtas2015TwoSiteJump}.

From the quantum state preparation perspective, the model exhibits competition between quantum jumps where the W-state is either an `attractive' ($L_1$, $L_2$) or `repulsive' ($L_3$, $L_{3'}$) dark state.  These properties are defined by the structure of the adjoint of the jump operators, $L^\dagger$.  Attractive dark states satisfy  $L^\dagger |W \rangle \neq 0$; this implies there is some attractive basin of states which flow into the W-state under $L$.  For repulsive dark states, $L|W\rangle = L^\dagger|W\rangle = 0$.  Here, states in the neighborhood of the W-state will never jump to the W-state (i.e. $\langle W|\rho(t)|W\rangle = \langle W|\rho(0)|W\rangle$), but will instead flow to some other steady state.  See \cite{supp} for further details.  



\textit{Dissipative W-State Preparation}---We now show that Lindbladian dynamics with $L_1$ and $L_2$ prepares the (generalized) W-state.  While this is not feasible in full generality---the generalized W-state falls outside the class of states preparable through local, purely dissipative Lindbladians \cite{Ticozzi2012DStatePrep}---various strategies have been explored to extend the set of preparable states \cite{Pechen2006control,Ticozzi2014DStatePrep,Scaramuzza2015switching,Grigoletto2022switching,Morales2024Steering} allowing for experimental W-state realizations \cite{Aron2016generalizedWState,Halati2017VortexLattice,Cole2021WStatePRep}.   


Here, the target state is prepared from a restricted subspace of allowed initial states.  Jump operators which commute with a symmetry of the target state may be used for dissipative preparation within the subspace of states that share that symmetry--Fig. \ref{fig: DPT Model}b.  Additional jump operators take other states into the target symmetry sector, broadening the space of initial states which prepare the target.


To prepare the W-state, we focus on subspaces of states with a fixed total $S_z$.  Within the 1-spin-up ($S_z=-\frac{N}{2}+1$) sector, the W-state is the only state with maximal total spin.  Applying the local jump operators $L_1$---which preserve $S_z$ while increasing total spin---therefore drives the one spin-up subspace to the W-state.  In other words, these jump operators locally lock the phase between sites, driving towards global phase coherence---an idea previously considered in \cite{Kraus2008DissipStatePrep,Diehl2008BECPrep} for similar jump operators in lattice ultracold bosonic atoms.  


Adding $L_2$---or any $L$ which takes a two spin-up state to a one spin-up state, including $L_3$---connects higher-$S_z$ sectors to the target $S_z=-\frac{N}{2}+1$ sector.  The full protocol, shown in Fig. \ref{fig: DPT Model}c,  combines $L_1$ to push each $S_z$-sector towards globally phase coherent states and $L_2$ to reduce $S_z$ until fewer than two spin-ups remain.  In this way, any initial state will converge to the W-state except for one - the vacuum (i.e. completely spin-down state $|\downarrow\downarrow...\downarrow\rangle$).  Like the W-state, the vacuum  is a dark state of $L_1$ and $L_2$.  Given the no-go theorem \cite{Ticozzi2012DStatePrep}, this is the closest possible procedure to fully dissipative preparation of the W-state; only one state is not driven by the dissipative attractor dynamics to the W-state, so a random initial state will converge to the W-state with probability 1 in the thermodynamic limit.  In supplementary material \cite{supp}, we analytically show preparation time for this protocol scales as $O(N^2)$. 

To prepare the generalized W-state, note that the generalized W-state is simply a local gauge transformation of the W-state, i.e. 

\begin{gather}
    |W_{\text{Gen}}\rangle = U_{\text{gauge}}|W\rangle = \prod_a \ee^{\ii \phi_a \sigma^+_a \sigma^-_a} |W\rangle
    \label{eq: Wsteady with phase}
\end{gather}
with $\sigma^\pm_a$ the spin raising/lowering operators and $\phi_a$ desired phase to imprint on site $a$.  A preparation strategy may then be constructed by applying this transformation to the jump operators, 
%
    $L_\ell \rightarrow U_{\text{gauge}} L_\ell U_{\text{gauge}}^\dagger
$, 
such that the dynamics under (\ref{eq: Lindbladian}) is preserved except to the new absorbing state $|W_{\text{Gen}}\rangle$.  

\textit{Absorbing State Phase Transition}---We now investigate what happens when $\gamma_3 \neq 0$.  Initially, we will focus on the all-to-all case for two main reasons.

First, the W-state is all-to-all symmetric and so leveraging an all-to-all preparation procedure might be expected to improve upon preparation time.  Indeed, the all-to-all model prepares the W-state in constant time as opposed to the polynomial scaling of the NN case.  This improvement, however, comes at a cost; the all-to-all procedure is fragile to $O(\frac{1}{N})$ error quantum jump rates, whereas the transition occurs for error jump rates $O(\frac{1}{Z})$ in the NN case.  

Second, any all-to-all Lindbladian (or general open quantum dynamics) may be represented by a set of generalized Dicke Operators, which act as closed operations within the space of generalized Dicke states \cite{Hartmann2016GenDicke}.  Specifically, any permutation symmetric density matrix may be represented as a linear combination of generalized Dicke states

\begin{equation}
    \rho = \sum_{q,q_z,\sigma_z} \alpha_{q,q_z,\sigma_z} {\cal D}_{q,q_z,\sigma_z}
    \label{eq: density of Dicke}
\end{equation}
where the generalized Dicke states ${\cal D}_{q,q_z,\sigma_z}$ are defined in terms of three quantum numbers $q$, $q_z$, and $\sigma_z$.  The number of ${\cal D}_{q,q_z,\sigma_z}$ is only $O(N^3)$, and therefore scales favorably compared to the exponential size of the full density matrix. 
Generalized Dicke states extend to open systems the more commonly used Dicke states, i.e. the set of permutation symmetric pure states whose utility was first introduced in the context of quantum optics \cite{Dicke1954DickeStates}.  We will focus here on the main results from the generalized Dicke analysis; calculation details may be found in supplementary material \cite{supp}.  

\begin{figure*}
    \centering
    \includegraphics[width=0.99\textwidth]{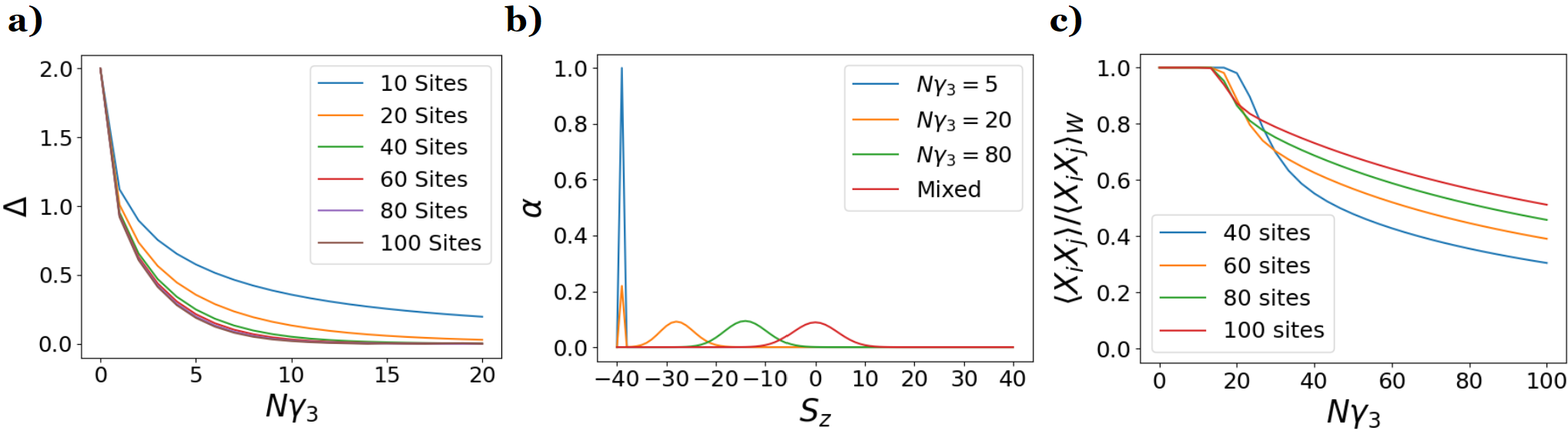}
    \caption{Absorbing state phase transition in the all-to-all system.  a) Gap to first non-zero eigenvalue.  Here, $\gamma_1=\gamma_2=1$ and numerics are done within the $\sigma_z=0$ subspace to access larger system sizes. b) Proportion of Steady state (with totally spin-up initial state) in each $S_z$ symmetry sector.  Numerics were done with $N=80$ and run for $t=1000$. c) Long range correlations in the steady state.  Transition point is consistent with closing of the gap in a).}
    \label{fig: DPT Numerics}
\end{figure*}

Using generalized Dicke operators, it is possible to see explicitly that the only two dark states of the system with $\gamma_3 = 0$ are the vacuum and the W-state. 
Furthermore, any initial state that has zero overlap with the vacuum will converge to the W-state.  The operators $L_3$, $L_{3'}$ on the other hand have 3 important steady states: the vacuum, the W-state, and the maximally mixed state (within the state space orthogonal to the two dark states).  Importantly, as remarked in our introduction of the model, both the vacuum and W-state are repulsive dark states.  Initial states with zero overlap in the vacuum or W-state will flow to the maximally mixed state.  

To numerically study the phase transition induced by these competing jump operators, we examine the gap $\Delta$ to the first non-zero eigenvalue.  The gap closes as $\gamma_3$ is increased (and $\gamma_1=\gamma_2=1$).  We plot the gap in Fig.~\ref{fig: DPT Numerics}a as a function of $N\gamma_3$.  The gap closing for such small values of $\gamma_3$---scaling with $\frac{1}{N}$ in the system size---implies that the W-state preparation strategy is fragile to the addition of $L_3$, $L_{3'}$.

A priori, this fragility may be surprising as $L_1$ and $L_{3'}$ jump from the same state (the singlet state) and act on the same number of site-pairs (similarly for $L_2$, $L_3$, and the $|1,1\rangle$ state).  There is, therefore, no $O(N)$ difference in the rate at which cohering vs. decohering quantum jumps occur.  This would seem to suggest a transition should occur when $\frac{\gamma_3}{\gamma_1} = O(1)$ not $O(\frac{1}{N})$.  


To see why this is not the case, it is helpful to look at the NN model and consider evolution of the expectation value $\langle P^{|1,0\rangle,|1,-1 \rangle} \rangle$ where $P^{|1,0\rangle,|1,-1 \rangle} = \frac{1}{N-1} \sum_{\langle a,b \rangle} \left( |1,0\rangle_{a,b} \langle 1,0|_{a,b} + |1,-1\rangle_{a,b} \langle 1,-1|_{a,b} \right)$.  Physically, $\langle P^{|1,0\rangle,|1,-1 \rangle} \rangle$ represents the expectation value that neighboring sites will be in either the $|1,0\rangle$ or $|1,-1\rangle$ state.  This is maximal for the W-state and serves as an order parameter for the transition.  As an example, consider evolution of this expectation value for any $\rho$ in the $S_z = -\frac{N}{2} + 1$ sector (see supplementary material for derivation with any $\rho$) given by

\begin{gather}
    \frac{d}{dt} \langle P^{|1,0\rangle,|1,-1 \rangle} \rangle = \left(\gamma_1 - \gamma_{3} \frac{Z-1}{2} \right) \langle  P^{|0,0\rangle} \rangle
    \label{eq: ddt of P}
\end{gather}
where $\langle P^{|0,0\rangle} \rangle$ is the expectation value that neighboring sites are in the $|0,0\rangle$ state. 
 Intuitively, the extra factor of $Z$ occurs for the decohering terms because---for any pair of sites $a,b$---the jump operators can decohere $a,b$ by acting on $a$ and any neighbor $c\neq b$ (or similarly, sites $b$ and a neighbor $c\neq a$).  However, to make $a,b$ coherent, $L_1$ must act directly on sites $a,b$.  In the all-to-all limit, $Z \rightarrow N$ and the decohering terms dominate by a factor of $N$, implying any non-zero $\gamma_{3}$ will ruin the W-state preparation procedure. 

We now investigate the properties of the steady state in the regime where the gap closes.  Here, the steady state will be a linear combination of the W-state and the eigendensity-matrix corresponding to the third $0$ eigenvalue.  The relative weights of the linear combination may in general depend on the initial state and the values of $\gamma_1,\gamma_2,\gamma_3$.  In figure \ref{fig: DPT Numerics}b, we compare steady states taking as our initial state the totally spin polarized system $|\uparrow\uparrow...\uparrow\rangle$.  The steady state is a mixture of density matrices with well defined total $S_z$ given by $\rho_{\text{steady}} = \sum_{S_z} \alpha_{S_z} \rho_{S_z}$ with $\sum_{S_z} \alpha_{S_z}=1$.  We plot $\alpha_{S_z}$ (given by $\alpha_{\frac{N}{2},S_z,0}$ in the generalized Dicke representation \eqref{eq: density of Dicke}) in Fig.~\ref{fig: DPT Numerics}b to show the relative weight of the steady state in each $S_z$ sector. 


When $\gamma_1,\gamma_2\ll\gamma_3$, the steady state (labeled as `mixed' in Fig. \ref{fig: DPT Numerics}b) is the maximally mixed state (minus the all spin-down and W-state).  From counting the number of states in each total $S_z$ sector, the distribution of coefficients for this `mixed' state is given by $\alpha^\text{Mixed}_{S_z} = \frac{\binom{N}{\frac{N}{2}+S_z}-\delta_{S_z,-\frac{N}{2}+1}}{2^N - 2}$,
where $\delta$ is the Kronecker delta and the $-2$ comes from the removal of the all spin-down and W-state.  As $\gamma_3$ decreases, the distribution shifts from the maximally mixed state to a distribution centered around smaller and smaller $S_z$.  Additionally, the weight of the W-state increases as $\gamma_3$ becomes small (with $\alpha_{-\frac{N}{2}+1}$ increasing towards 1).  In the regime where the gap opens, the steady state becomes the W-state.

 As a measure of coherence in the system, we consider the two-point correlator $\langle X_a X_b \rangle$ for Pauli $X_a$ at site $a$.  For the W-state, we have $\langle X_a X_b \rangle_W = \frac{2}{N}$.  In Fig. \ref{fig: DPT Numerics}c, we compare $\langle X_a X_b \rangle$ for the steady state of our model to that of the W-state.  The ratio $\frac{\langle X_a X_b \rangle}{\langle X_a X_b \rangle_W}$ departs from $1$ at the same value of $\gamma_3$ where the gap closes (compare to Fig. \ref{fig: DPT Numerics}a).  Coherence still persists, however, after the transition. 

The steady state, therefore, does not just converge to the steady state of $L_3$ (the maximally mixed state) after the transition.  Instead, the steady state is a mixture of states with coherence less than, but comparable to, that of the W-state.  It is only in the limit $\gamma_3 \gg \gamma_1,\gamma_2$ that this coherence approaches $0$ and the steady state becomes maximally mixed.

\textit{NN Results and Criticality}---We now contrast the above all-to-all results with the NN version of the model, as well as discuss critical behavior.  As previously discussed, the transition in the NN model occurs when $\frac{\gamma_3}{\gamma_1} = O(1)$.  Furthermore, unlike the all-to-all transition which goes from gapped to gapless, the NN transition is from  a polynomial gap to an exponential gap in system size.  The replacement of the fixed gap (for all-to-all) with a polynomial gap is because the local jump operators in the NN model need time at least on order of the system size to generate coherence across the entire system.  In the mixed phase, the NN model also exhibits coherence, but in this case the coherence decays exponentially in $m$ for $\langle X_a X_{a + m} \rangle$ again due to locality.  Numerics for small system sizes supporting these results are provided in the supplementary material. 

To analyze behavior at the critical point, we will focus on the dynamical critical exponent $z$ given by the gap $\Delta = O(N^{-z})$.  For 1D systems in the DP universality class, $z\approx 1.5807$ \cite{HinrichsenReactionDiffusion}.  However, for the NN model, we must have $z\geq2$.  This is because the gap in the $\gamma_3=0$ limit (i.e. in the ordered phase) already scales as $\Delta=O(N^{-2})$ \cite{supp}, and the addition of $\gamma_3$ terms may only increase the dynamical exponent.  Small system size numerics yield the critical value $z\approx2.7$ at the transition into the disordered phase.  Thus, the critical behavior of the NN model is inconsistent with DP universality.  The all-to-all model, on the other hand, did not show numerical inconsistencies with DP \cite{supp}.

\textit{Conclusions}---We have investigated a purely dissipative quantum reaction-diffusion model which contains competing terms that act to either cohere or decohere the system.  The cohering terms may be viewed as a dissipative quantum state preparation protocol for the generalized W-state, while the decohering terms act as errors to the preparation procedure.  This transition---from a long range coherent absorbing state to a mixed phase with short range coherence---leads to critical behavior which falls outside the directed percolation universality class typical for absorbing state phase transitions.  This distinct behavior is due to the dynamcis of coherence spreading throughout the system at criticality.  We, furthermore, showed that while the all-to-all model can quickly prepare the W-state, the preparation protocol is fragile to errors: the critical error rate scales as the inverse of the system size (whereas the nearest neighbour model is robust to weak error rates).  This leads to a tradeoff between nearest neighbor models which prepare slowly but are robust to errors, and all-to-all models which prepare quickly but are fragile to errors.   
Investigating to what extent all-to-all dissipative state preparation procedures more generally are fragile to errors is an interesting avenue for future work.   

While we have shown that coherence spreading at the transition can lead to non-DP critical behavior, many open questions remain surrounding the nature of this transition.  For example, we have established distinction from DP through a lower bound on the dynamical critical exponent, but it would be interesting to establish accurate estimates of all critical exponents through numerical or analytical methods.  Furthermore, the universality of such critical exponents is also unclear as---for any dissipative preparation procedure with an absorbing state transition---the dynamical critical exponent must be lower bounded by the preparation time without errors.  This suggests the possibility of distinct critical behavior depending on the dissipative preparation protocol used. 

It is also an open question how much of a role the vacuum plays in our system.  While it is disconnected from the dynamics, regions which are indistinguishable locally from the vacuum must also be dark and therefore will exhibit slowed dynamics.  Such dynamics is typical in kinetically constrained models, e.g. \cite{Olmos2012KCM,Causer2025KCM}.  A full analysis of how this physics affects $z$ and the dynamics more broadly we leave to future work.  On the other hand, it is also possible to isolate the W-state as the sole dark state through the addition of coherent terms \cite{Ticozzi2014DStatePrep}.  Though more rigorous analysis is needed, preliminary small system size numerical results when this coherent term is added are also consistent with critical $z>2$ \cite{supp}.



\begin{acknowledgments}
This work was supported by a Simons Investigator Award
(Grant No. 511029) and the Engineering and Physical Sciences Research Council (Grants No. EP/V062654/1 and No.
EP/Y01510X/1).
\end{acknowledgments}
\bibliography{main.bib}

\onecolumngrid
\appendix

\section{Attractive and Repulsive Dark States}
\label{sec: attractive and repulsive explanation}

\begin{figure}[h]
    \centering
    \includegraphics[width=0.55\linewidth]{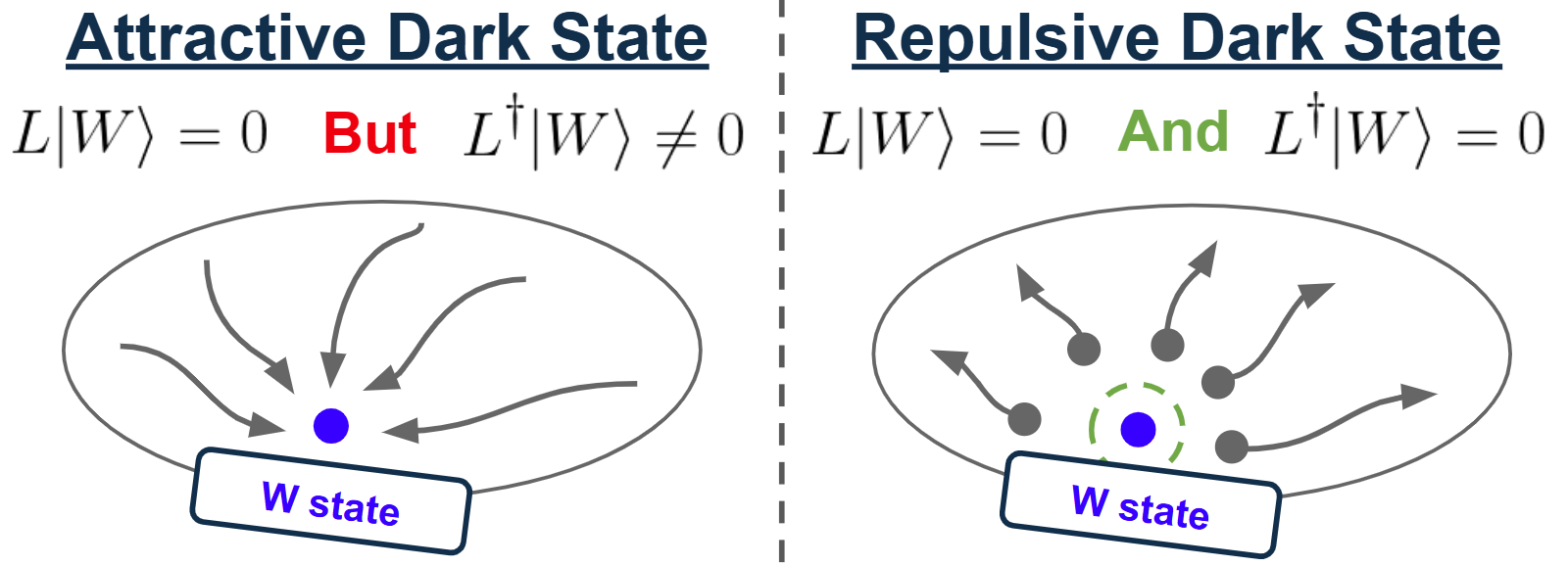}
    \caption{Visual representation of attractive vs. repulsive dark states.  Points and arrows correspond to states within Hilbert space and the action of the jump operator $L$ on those states respectively.}
    \label{fig: dark states}
\end{figure}

Figure \ref{fig: dark states} provides a visual representation of the difference between attractive and repulsive dark states.  The points correspond to states and arrows correspond to the action of a jump operator $L$ on those states. 
 In both the attractive and repulsive cases of Fig. \ref{fig: dark states}, the W-state is a dark state of $L$---this corresponds to no arrows coming out of the W-state vertex, as dissipative evolution under $L$ acts as the identity on the W-state.  To see whether Lindblad evolution under $L$ will take other states \textit{into} the W-state, it is most convenient to look at the action of $L^\dagger$.  If the action of $L^\dagger$ can evolve the system \textit{out} of the W-state, then this implies the action of $L$ takes some other state \textit{into} the W-state.  It is for this reason that, when $L|W\rangle = L^\dagger|W\rangle = 0$, we refer to the W-state as a \textit{repulsive} dark state since---for evolution of $\rho(t)$ under a Lindbladian with jump operator $L$---no states can evolve into or out of the W-state, i.e. $\langle W|\rho(t)|W\rangle = \langle W|\rho(0)|W\rangle$.  If the W-state is instead an \textit{attractive} dark state of $L$, this implies that $\langle W|\rho(t)|W\rangle \geq \langle W|\rho(0)|W\rangle$ (which is a necessary, but not sufficient, condition for the Lindbladian to be a dissipative W-state preparation procedure, i.e.  $\lim_{t\rightarrow \infty} \langle W|\rho(t)|W\rangle = 1$).
  
Another way to understand our definition of attractive and repulsive is to look at the Lindblad operator for a quantum jump $L$ in the doubled space ${\cal L}_L \equiv L \otimes L^* - \frac{1}{2} \left[L^\dagger L \otimes I + I \otimes (L^\dagger L)^T \right]$.  Note that if the W-state is a repulsive dark state, then ${\cal L}_{L}^\dagger|W\rangle |W\rangle = {\cal L}_{L}|W\rangle |W\rangle = 0$ and we have that ${\cal L}_L = P_{\perp |W\rangle|W\rangle}{\cal L}_{L} P_{\perp |W\rangle|W\rangle}$---where $P_{\perp |W\rangle|W\rangle}$ is the projector onto the space of (vectorized) density matrices orthogonal to the pure W-state---i.e. the dynamics of ${\cal L}_L$ does not take any states into or out of the W-state.

\section{Generalized Dicke States and Operators}

We briefly review here generalized Dicke states and operators.  See \cite{Hartmann2016GenDicke} for further details and relations.  Consider dynamics under Lindblad operators (or, more generally, Kraus operators) which act symmetrically on a set of $N$ sites.  As first noted (independently) in \cite{Chase2008GenDicke,Hartmann2016GenDicke}, both analytical and numerical investigations into such scenarios may be dramatically simplified by leveraging the fact that the set of Lindblad operators of this type is closed under $SU(4)$ transformations.  As we will see below, the number of basis states for the space of symmetric density matrices scales only polynomially in $N$, providing an exponential improvement over the $2^N \times 2^N$ scaling of the full density matrix.  In analogy with Dicke states -- the set of fully permutation symmetric pure states (closed under SU(2) operations instead of SU(4)) that form the appropriate basis for Hamiltonians which act symmetrically on all sites -- the basis of fully symmetric density matrices are referred to as generalized Dicke states and the generators of $SU(4)$ which act on them are referred to as generalized Dicke operators.

Specifically, the set of all permutation symmetric Lindblad operators may be generated by:
\begin{gather}\label{eq: generalized Dicke Operators}
\begin{aligned}
    &{\cal Q}_\pm [\rho] \coloneqq \sum_{j=1}^{N} \sigma_j^{\pm} \rho \sigma_j^{\mp}; \; \; {\cal Q}_z [\rho] \coloneqq \frac{1}{4} \sum_{j=1}^{N} \left(\sigma_j^z \rho + \rho \sigma_j^z \right) \\
    &\Sigma_\pm [\rho] \coloneqq \sum_{j=1}^{N} \sigma_j^{\pm} \rho \sigma_j^{\pm}; \; \; \Sigma_z [\rho] \coloneqq \frac{1}{4} \sum_{j=1}^{N} \left(\sigma_j^z \rho - \rho \sigma_j^z \right) \\
    &{\cal M}_\pm [\rho] \coloneqq \sum_{j=1}^{N} \sigma_j^{\pm} \rho \frac{1 + \sigma_j^z}{2}; \; \; {\cal M}_z [\rho] \coloneqq \frac{1}{2} \sum_{j=1}^{N} \sigma_j^z \rho \frac{1 + \sigma_j^z}{2} \\
    &{\cal N}_\pm [\rho] \coloneqq \sum_{j=1}^{N} \sigma_j^{\pm} \rho \frac{1 - \sigma_j^z}{2}; \; \; {\cal N}_z [\rho] \coloneqq \frac{1}{2} \sum_{j=1}^{N} \sigma_j^z \rho \frac{1 - \sigma_j^z}{2} \\
    &{\cal U}_\pm [\rho] \coloneqq \sum_{j=1}^{N} \frac{1 + \sigma_j^z}{2} \rho \sigma_j^{\mp}; \; \; {\cal U}_z [\rho] \coloneqq \frac{1}{2} \sum_{j=1}^{N} \frac{1 + \sigma_j^z}{2} \rho \sigma_j^z \\
    &{\cal V}_\pm [\rho] \coloneqq \sum_{j=1}^{N} \frac{1 - \sigma_j^z}{2} \rho \sigma_j^{\mp}; \; \; {\cal V}_z [\rho] \coloneqq \frac{1}{2} \sum_{j=1}^{N} \frac{1 - \sigma_j^z}{2} \rho \sigma_j^z 
\end{aligned}
\end{gather}
Of the 18 generalized Dicke operators listed above, 15 are linearly independent and correspond to the generators of $SU(4)$. 

The generalized Dicke states are fundamentally represented in terms of the single site density operators

\begin{gather}
\begin{aligned}
    &u = |1\rangle\langle1|, \; d = |0\rangle\langle0| \\
    &s = |1\rangle\langle0|, \; c = |0\rangle\langle1|
\end{aligned}
\end{gather}
where $u,d,s,c$, i.e. up, down, strange, and charm, are a reference to the $SU(4)$ four-flavor quark model \cite{Georgi2000LieParticle}.
 Specifically, the generalized Dicke states are given by

\begin{gather}
    {\cal D}_{q,q_z,\sigma_z} = {\cal S}\left(u^\alpha d^\beta s^\gamma c^\delta \right)
    \label{eq: generalized Dicke states}
\end{gather}
where ${\cal S}$ is the symmetrizer and $\alpha + \beta + \gamma + \delta = N$.  The quantum numbers $q$, $q_z$, $\sigma_z$ are given in terms of $\alpha$, $\beta$, $\gamma$, $\delta$ by

\begin{gather}
    q = \frac{\alpha + \beta}{2},\\
    q_z = \frac{\alpha - \beta}{2},\\
    \sigma_z = \frac{\gamma - \delta}{2}.
\end{gather}
The range of possible $q,q_z,\sigma_z$ are $q=0,\frac{1}{2},...,\frac{N}{2}$; $q_z=-q,-q+1,...,q$; and $\sigma_z=-\frac{N}{2}+q,-\frac{N}{2}+q+1,...,\frac{N}{2}-q$.  By simple counting arguments, this implies there are a total of $\frac{1}{6}(N+1)(N+2)(N+3)=O(N^3)$ generalized Dicke states.  

It is possible to denote the action of the generalized Dicke operators \eqref{eq: generalized Dicke Operators} onto the states \eqref{eq: generalized Dicke states}.  Namely, 
\begin{equation}
\begin{tabular}{c | c | c }\centering
    $\text{Operator} \;$ & $\; (q,q_z,\sigma_z) \;$ & $\; +,-$\\ \hline 
    ${\cal Q}_\pm$ & $(0,\pm 1,0)$  & $\beta,\alpha$\\
    ${\Sigma}_\pm$ & $(0,0,\pm 1)$  & $\delta,\gamma$\\
    ${\cal M}_\pm$ & $\;(\mp \frac{1}{2},\pm \frac{1}{2},\pm \frac{1}{2})\;$  & $\beta,\gamma$\\
    ${\cal N}_\pm$ & $(\pm \frac{1}{2},\pm \frac{1}{2},\pm \frac{1}{2})$  & $\delta,\alpha$\\
    ${\cal U}_\pm$ & $(\mp \frac{1}{2},\pm \frac{1}{2},\mp \frac{1}{2})$  & $\beta,\delta$\\
    ${\cal V}_\pm$ & $(\pm \frac{1}{2},\pm \frac{1}{2},\mp \frac{1}{2})$  & $\gamma,\alpha$\\
\end{tabular}
\quad
\begin{tabular}{c|c}
    Operator \; & \; Coeff.  \\
    \hline
     ${\cal Q}_z$ & $\frac{\beta-\alpha}{2} = -q_z$ \\ 
     ${\Sigma}_z$ & $\frac{\delta-\gamma}{2}=-\sigma_z$ \\ 
     ${\cal M}_z$ & $\frac{\beta-\gamma}{2}$ \\ 
     ${\cal N}_z$ & $\frac{\delta-\alpha}{2}$ \\ 
     ${\cal U}_z$ & $\frac{\beta-\delta}{2}$ \\ 
     ${\cal V}_z$ & $\frac{\gamma-\alpha}{2}$ \\ 
\end{tabular}
\end{equation}
where the table above denotes how each operator alters the quantum numbers and what coefficient it adds to the generalized Dicke state.  For example,

\begin{gather}
    {\cal Q}_+ {\cal D}_{q,q_z,\sigma_z} = \beta {\cal D}_{q,q_z + 1,\sigma_z}, \\
    {\cal Q}_z {\cal D}_{q,q_z,\sigma_z} = -q_z {\cal D}_{q,q_z,\sigma_z}.
\end{gather}

Due to the algebra above, it will often be useful to represent permutation symmetric Lindbladians as transformations on a 3 dimensional lattice with each dimension corresponding to a quantum number of the generalized Dicke states.  Each site is labeled by a generalized Dicke state

\begin{gather}
    {\cal D}_{q,q_z,\sigma_z} \equiv |q,q_z,\sigma_z\rangle,
\end{gather}
and a general symmetric Lindbladian ${\cal L}$ will have matrix elements of the form

\begin{gather}
    {\cal L} = \sum_{\substack{q,q_z,\sigma_z \\ q',q_z',\sigma_z'}} {\cal L}(q',q_z',\sigma_z';q,q_z,\sigma_z) |q',q_z',\sigma_z'\rangle \langle q,q_z,\sigma_z|.
    \label{eq: Lindblad Representation}
\end{gather}
In general, a density matrix will take the form of a linear combination of generalized Dicke states

\begin{gather}
    \rho = \sum_{q,q_z,\sigma_z} \alpha_{q,q_z,\sigma_z} |q,q_z,\sigma_z \rangle .
    \label{eq: rho as Dicke}
\end{gather}
We will often omit quantum numbers on the coefficients for density matrices confined to a subspace with a fixed quantum number.  For example, if the evolution of a density matrix is confined within the $\sigma_z=0$ sector, we will write its coefficients as $\alpha_{q,q_z}$.  We note that the $\alpha_{q,q_z,\sigma_z}$ are constrained by the condition $\Tr{\rho} = 1$, in particular

\begin{gather}
    \sum_{q_z} \alpha_{\frac{N}{2},q_z,0} = 1.
\end{gather}

\section{All-to-All Model with Generalized Dicke Operators}

We here write the Lindbladian for the all-to-all version of our model in terms of generalized Dicke operators.  We subdivide the Lindbladian into terms

\begin{gather}
    {\cal L} = {\cal L}_1 + {\cal L}_2 + {\cal L}_3 + {\cal L}_{3'}
\end{gather}
with each ${\cal L}_\alpha$ denoting the part of the Lindblad superoperator which corresponds to the set of jump operators \eqref{eq: L1}, \eqref{eq: L2}, \eqref{eq: L3}, or \eqref{eq: L3'} in the main text (with all pairs $i,j$ instead of nearest neighbors $<i,j>$).  In other words, we define

\begin{gather}
    {\cal L}_\alpha \left[ \rho \right] = \sum_{L \in L_\alpha} \gamma_\alpha \left( L \rho L^\dagger - \frac{1}{2} \left\{L^\dagger L,\rho \right\}\right).  \label{eq: Lop alpha}
\end{gather}
where each $L \in L_\alpha$ corresponds to the jump operator of form $L_\alpha$ acting on a given site pair $i,j$.

In terms of generalized Dicke operators, the operators ${\cal L}_\alpha$ may be written

\begin{gather}
    {\cal L}_1 = \gamma_1 \left\{ {\cal M}_+ {\cal N}_- + {\cal U}_+ {\cal V}_- - 2 {\cal S} {\cal C} \right\} \label{eq: Lop 1}\\
    {\cal L}_2 = \gamma_2 \left\{{\cal N}_- {\cal V}_- +  {\cal Q}_-\left[\frac{N}{4} - {\cal Q}_z + \frac{1}{2}({\cal M}_z - {\cal N}_z) -1\right] -\left(\frac{N}{2} - {\cal Q}_z\right)\left(\frac{N}{2} - {\cal Q}_z - 1\right) -\Sigma_z^2\right\} \label{eq: Lop 2}\\
    \begin{aligned}
    {\cal L}_3 + {\cal L}_{3'} = \gamma_3 \left\{ \left({\cal Q}_- + {\cal Q}_+ \right)\left[\frac{N}{4} - {\cal Q}_z + \frac{1}{2}({\cal M}_z - {\cal N}_z)\right] - {\cal Q}_- - N\left(\frac{N}{2} - {\cal Q}_z \right) \right. \\
    \left. -{\cal N}_+ {\cal V}_+ -{\cal N}_- {\cal V}_- +\frac{1}{2} \left[ \left({\cal M}_+ + {\cal N}_+\right)\left({\cal M}_- + {\cal N}_-\right) + \left({\cal U}_+ + {\cal V}_+\right)\left({\cal U}_- + {\cal V}_-\right) \right] \right\}. \label{eq: Lop 3}
    \end{aligned}
\end{gather}
where ${\cal S}$ and ${\cal C}$ are given by

\begin{gather}
    {\cal S} = \frac{N}{4} + {\cal N}_z - \frac{1}{2} {\cal Q}_z - \frac{3}{2} \Sigma_z \\
    {\cal C} = \frac{N}{4} + {\cal N}_z - \frac{1}{2} {\cal Q}_z + \frac{1}{2} \Sigma_z
\end{gather}
and correspond to the the number operators for the number of $s$ and $c$ states in the generalized Dicke state, i.e. 

\begin{gather}
    {\cal S} {\cal D}_{q,q_z,\sigma_z} = \left(\frac{N}{2} - q + \sigma_z \right){\cal D}_{q,q_z,\sigma_z} = \gamma {\cal D}_{q,q_z,\sigma_z} \\
    {\cal C} {\cal D}_{q,q_z,\sigma_z} = \left(\frac{N}{2} - q - \sigma_z \right){\cal D}_{q,q_z,\sigma_z} = \delta {\cal D}_{q,q_z,\sigma_z}.
\end{gather}

Equations \eqref{eq: Lop 1}, \eqref{eq: Lop 2}, and \eqref{eq: Lop 3} were found by writing the jump operators in terms of Pauli operators

\begin{gather}
    |1,0\rangle_{i,j}\langle 0,0|_{i,j} = \frac{\sigma_i^z - \sigma_j^z - i \sigma_i^y \sigma_j^x + i \sigma_i^x \sigma_j^y}{4},\\
    |1,0\rangle_{i,j} \langle 1,1|_{i,j} = \frac{1}{\sqrt{2}} \left[\sigma_i^- \left(\frac{I - \sigma_j^z}{2} \right) +  \left(\frac{I - \sigma_i^z}{2} \right) \sigma_j^-  \right] ,\\
    |0,0\rangle_{i,j} \langle 1,1|_{i,j} = \frac{1}{\sqrt{2}} \left[\sigma_i^- \left(\frac{I - \sigma_j^z}{2} \right) - \left(\frac{I - \sigma_i^z}{2} \right) \sigma_j^- \right], \\
    |1,1\rangle_{i,j} \langle 0,0|_{i,j} = \frac{1}{\sqrt{2}} \left[\sigma_i^+ \left(\frac{I - \sigma_j^z}{2} \right)  -  \left(\frac{I - \sigma_i^z}{2} \right) \sigma_j^+ \right]. 
\end{gather}
in Eq. \eqref{eq: Lop alpha} and using $\sum_{i,j \; s.t. \; i\neq j} = \sum_{i,j} - \sum_{i \; s.t. \; i=j}$.  Above, we have written 2-site states in terms of total spin eigenstates $|s,m_s\rangle$ given by

\begin{gather}
\begin{aligned}
    &|1,1\rangle_{ij} = |\uparrow \uparrow \rangle_{ij}, \;\; |1,0\rangle_{ij} = \frac{1}{\sqrt{2}}(|\downarrow \uparrow\rangle_{ij} + |\uparrow\downarrow\rangle_{ij}),\\
    &|1,-1\rangle_{ij} = |\downarrow\downarrow\rangle_{ij}, \;\; |0,0\rangle_{ij} = \frac{1}{\sqrt{2}}(|\downarrow\uparrow\rangle_{ij} - |\uparrow\downarrow\rangle_{ij}).\\
\end{aligned}
\end{gather}

\section{W-State Preparation}
\label{Section: gDicke W state Prep}

In this section, we show that ${\cal L}_1 + {\cal L}_2$ together prepare the W-state.  Let us begin by solving for the steady state of ${\cal L}_1$.  By acting with ${\cal L}_1$ on the state ${\cal D}_{q,q_z,\sigma_z}$, the matrix elements of ${\cal L}_1$ in the representation \eqref{eq: Lindblad Representation} may be found to be 

\begin{gather}
\begin{aligned}
    &{\cal L}_{1} (q-1;q) = 2(q+q_z)(q-q_z) \\
    &{\cal L}_{1} (q;q) = -2\left(\frac{N}{2} - q + \sigma_z \right)\left(\frac{N}{2} - q - \sigma_z \right)  
\end{aligned}
\label{eq: LWn GDicke action on GDicke states}
\end{gather}
where we have denoted terms which leave $q_z,\sigma_z$ invariant as ${\cal L}_{1} (q-1;q) \equiv {\cal L}_{1} (q-1,q_z,\sigma_z;q,q_z,\sigma_z)$.
We will refer to the first term above, which takes $q\rightarrow q-1$, as the ``hopping'' term and the second term above, which leaves $q$,$q_z$,$\sigma_z$ invariant, as the ``on-site'' term.  Note, as expected, \eqref{eq: LWn GDicke action on GDicke states} does not alter $q_z$ and thus there will be a different, independent solution for each symmetry sector $q_z$.  Similarly, $\sigma_z$ is also invariant under the action of the Lindbladian.  We therefore focus on the evolution of $q$ to find the steady state of the dynamics.  The steady state, for any given sector of $q_z,\sigma_z$, will be some linear combination of $q$ generalized Dicke states

\begin{gather}
    {\cal D}_{\text{steady}} = \sum_q \alpha_q {\cal D}_{q,q_z,\sigma_z}.
    \label{eq:GDicke Steady}
\end{gather}
The steady state condition ${\cal L}_{1}{\cal D}_{\text{steady}}=0$ then implies that the on-site term times $\alpha_q$ must be equal and opposite to the hopping term times $\alpha_{q+1}$ for all $q$.  

Instead of solving directly for $\alpha_q$, we solve for a related quantity, $\alpha_q^{(N)}$ defined below, which will help make the interpretation of the result more transparent.  We remark that each $\alpha_{q,q_z,\sigma_z}$ corresponds to the sum of all the elements within the $q,q_z,\sigma_z$ sector of the full density matrix [see \eqref{eq: rho as Dicke}].  Comparing different $\alpha_q$ is thus obscured by the fact that the different $q,q_z,\sigma_z$ sectors vary in size.  For a more direct comparison we define $\alpha_q^{(N)}$ as the weight of a single matrix element within the $q,q_z,\sigma_z$ sector of the density matrix. In particular, let ${\cal N}(q,q_z,\sigma_z)$ be the number of density matrix elements within the $q,q_z,\sigma_z$ sector, i.e.

\begin{gather}
    {\cal N}(q,q_z,\sigma_z) = \binom{N}{\alpha} \binom{N-\alpha}{\beta} \binom{N-\alpha-\beta}{\gamma} \nonumber \\
    = \binom{N}{q+q_z} \binom{N-q-q_z}{q-q_z} \binom{N-2q}{\frac{N}{2} - q + \sigma_z}.
\end{gather}
where the above is calculated using \eqref{eq: generalized Dicke states} and basic countig arguments.  

Therefore, we may now define
\begin{gather}
    \alpha_q = {\cal N}(q,q_z,\sigma_z) \alpha_q^{(N)},
\end{gather}

 It is possible to work directly in a basis which yields $\alpha_q^{(N)}$ instead of $\alpha_q$ by renormalizing the generalized Dicke states 
 
\begin{gather}
    |q,q_z,\sigma_z\rangle^{(N)} = {\cal N}(q,q_z,\sigma_z) |q,q_z,\sigma_z\rangle, \\
    \langle q,q_z,\sigma_z|^{(N)} = \frac{\langle q,q_z,\sigma_z|}{{\cal N}(q,q_z,\sigma_z)}.
\end{gather}
Writing any ${\cal L}$ in this basis yields

\begin{gather}
    {\cal L} = \sum_{\substack{q,q_z,\sigma_z \\ q',q_z',\sigma_z'}} {\cal L}^{(N)}(q',q_z',\sigma_z';q,q_z,\sigma_z) |q',q_z',\sigma_z'\rangle^{(N)} \langle q,q_z,\sigma_z|^{(N)} \\
    {\cal L}^{(N)}(q',q_z',\sigma_z';q,q_z,\sigma_z) = {\cal L}(q',q_z',\sigma_z';q,q_z,\sigma_z) \frac{{\cal N}(q,q_z,\sigma_z)}{{\cal N}(q',q_z',\sigma_z')}.
\end{gather}

 Using that $\frac{{\cal N}(q,q_z,\sigma_z)}{{\cal N}(q-1,q_z,\sigma_z)} = \frac{\left(\frac{N}{2} - q + \sigma_z + 1 \right)\left(\frac{N}{2} - q - \sigma_z + 1\right)}{(q+q_z)(q-q_z)}$, rewriting \eqref{eq: LWn GDicke action on GDicke states} in the renormalized basis gives

\begin{gather}
\begin{aligned}
    &{\cal L}_{1}^{(N)} (q;q+1) = 2\left(\frac{N}{2} - q + \sigma_z \right)\left(\frac{N}{2} - q - \sigma_z \right), \\
    &{\cal L}_{1}^{(N)} (q;q) = -2\left(\frac{N}{2} - q + \sigma_z \right)\left(\frac{N}{2} - q - \sigma_z \right).
\end{aligned}
\label{eq: LWn GDicke action on GDicke states renormalized}
\end{gather}
In the renormalized basis, the hopping term is equal and opposite to the onsite term.  Letting ${\cal D}_{\text{steady}}^{(N)} = \sum_q \alpha_q^{(N)} {\cal D}_{q,q_z,\sigma_z}^{(N)}$, the steady state condition implies

\begin{gather}
\begin{aligned}
    &{\cal L}_{1} {\cal D}_{\text{steady}}^{(N)} = 0 \\
    \implies \alpha_q^{(N)} {\cal L}_{1}^{(N)} &(q;q) + {\alpha}_{q+1}^{(N)} {\cal L}_{1}^{(N)} (q;q+1) = 0 \;\; \forall \; q \\
    &\implies {\alpha}_{q+1}^{(N)} = {\alpha}_{q}^{(N)}
\end{aligned}
\end{gather}

Note that, since the generalized Dicke states which correspond to the diagonal of the full density matrix all have $q=\frac{N}{2}$ and since $\Tr \rho = 1$, we have that ${\alpha}_{\frac{N}{2}}^{(N)} = \frac{1}{{\cal N}(\frac{N}{2},q_z,0)} =\frac{1}{\binom{N}{\frac{N}{2}+q_z}}$. Therefore, 

\begin{gather}
    {\alpha}_{q}^{(N)} = \frac{1}{\binom{N}{\frac{N}{2}+q_z}}.
\end{gather}

For the case $q_z=-\frac{N}{2} + 1$ and $\sigma_z=0$, this is precisely given by the $W$ state

\begin{gather}
    \rho_{W} = |W\rangle \langle W|.
\end{gather}

More generally, the steady states corresponding to any $\sigma_z=0$ sector are given by the pure Dicke state corresponding to the given $q_z$.  The $\sigma_z\neq0$ terms correspond to superpositions between the different particle number sectors.  We note that all such $\sigma_z\neq 0$ generalized Dicke states have zero trace, and so must be accompanied (in linear combination) by $\sigma_z=0$ generalized Dicke states to be a steady state.  Furthermore, the hermiticity of the density matrix requires that the weight of ${\cal D}_{q,q_z,\sigma_z}$ and ${\cal D}_{q,q_z,-\sigma_z}$ must be related by a complex conjugate in the steady state. 

As a technical note, the renormalized basis breaks down at the boundaries of the 3D lattice.  Before renormalizing, any physical matrix element ${\cal L}(q',q_z',\sigma_z';q,q_z,\sigma_z)$ will be $0$ if it maps a site on the lattice to a $q',q_z',\sigma_z'$ which falls outside the range of possible values for the quantum numbers.  However, ${\cal N}(q',q_z',\sigma_z')$ is undefined for any choice of $q',q_z',\sigma_z'$ which is not within the allowed ranges.  This leaves ${\cal L}^{(N)}(q',q_z',\sigma_z';q,q_z,\sigma_z)$ ill-defined.  For example, at lattice site $|\frac{N}{2},-\frac{N}{2},0\rangle$, the non-renormalized matrix element ${\cal L}_1(\frac{N}{2}-1,\frac{N}{2}) = 0$ as this would make $q_z$ lie outside its allowed range of $-q$ to $q$.  However, the corresponding renormalized matrix element ${\cal L}_1^{(N)} (\frac{N}{2}-1,\frac{N}{2}) = 2 \neq 0$.  It is therefore necessary to reset all such boundary terms to $0$ after renormalizing.         

Let us now add ${\cal L}_2$ to the dynamics.  First consider the $\sigma_z = 0$ sector.  Here, the steady state is expanded in terms of generalized Dicke states as

\begin{gather}
    {\cal D}_{\text{steady}}^{(N)} = \sum_q {\alpha}_{q,q_z}^{(N)} {|q,q_z,\sigma_z\rangle}^{(N)}
\end{gather}
with the coefficients given by the system of equations $\left({\cal L}_{1} + {\cal L}_{2} \right) {\cal D}_{\text{steady}}^{(N)} = 0$.  Specifically, we have

\begin{gather}
    {\alpha}_{q,q_z}^{(N)} \left({\cal L}_{1}^{(N)}(q,q_z;q,q_z) + {\cal L}_{2}^{(N)}(q,q_z;q,q_z) \right) + {\alpha}_{q+1,q_z}^{(N)} {\cal L}_{1}^{(N)} (q,q_z;q+1,q_z) \nonumber \\
    + {\alpha}_{q,q_z + 1}^{(N)} {\cal L}_{2}^{(N)} (q,q_z;q,q_z + 1) + {\alpha}_{q+1,q_z + 1}^{(N)} {\cal L}_{2}^{(N)}(q,q_z;q+1,q_z + 1) = 0  \;\; \forall \; q,q_z
    \label{eq: steady SOE}
\end{gather}

In the system of equations above, the term ${\alpha}_{\frac{N}{2}^{(N)},\frac{N}{2}}$ only appears in the equation ${\alpha}_{\frac{N}{2},\frac{N}{2}}^{(N)} \left({\cal L}_{1}^{(N)}(\frac{N}{2},\frac{N}{2};\frac{N}{2},\frac{N}{2}) + {\cal L}_{2}^{(N)}(\frac{N}{2},\frac{N}{2};\frac{N}{2},\frac{N}{2}) \right) = - \gamma_2 N (N-1) {\alpha}_{\frac{N}{2},\frac{N}{2}}^{(N)} = 0$.  This implies ${\alpha}_{\frac{N}{2},\frac{N}{2}}^{(N)} = 0$.  By induction, all terms ${\alpha}_{q,q_z}^{(N)} = 0$ when $q_z > -\frac{N}{2} + 1$.  This may be seen by using \eqref{eq: steady SOE} and taking the inductive step $q \rightarrow q-1$ if $q > q_z$ and $q \rightarrow \frac{N}{2}$, $q_z \rightarrow q_z - 1$ otherwise.  In each inductive step, we are left with ${\alpha}_{q,q_z}^{(N)} \left({\cal L}_{1}^{(N)}(q,q_z;q,q_z) + {\cal L}_{2}^{(N)}(q,q_z;q,q_z) \right)  = 0$ as all ${\alpha}_{q,q_z}^{(N)}$ for larger $q,q_z$ are $0$ by the inductive hypothesis.  When $q_z > -\frac{N}{2} + 1$, $\left({\cal L}_{1}^{(N)}(q,q_z;q,q_z) + {\cal L}_{2}^{(N)}(q,q_z;q,q_z) \right) \neq 0$ which implies all ${\alpha}_{q,q_z}^{(N)} = 0$ for $q_z > -\frac{N}{2} + 1$.  When $q_z \leq -\frac{N}{2} + 1$, all ${\cal L}_2^{(N)}$ matrix elements are $0$, thus the longtime dynamics is completely determined by ${\cal L}_1$ in these sectors.  As we saw previously in this section, ${\cal L}_1$ prepares a W-state for $q_z=-\frac{N}{2} + 1$ and the vacuum for $q_z=-\frac{N}{2}$.  Since there is no hopping term which connects the $q_z = -\frac{N}{2}$ sector to the rest of the lattice, then any initial state starting outside this sector (and with $\sigma_z=0$) will converge to the W-state.  Outside of the $\sigma_z=0$ sector, we can apply a similar inductive argument, except this time there are no states where the on-site term is $0$ (except for $\sigma_z=\pm\frac{1}{2}$ and $q_z=-\frac{N}{2}+\frac{1}{2}$ which correspond to states in a superposition between the vacuum and the 1 particle sector).  This means, for any initial state which does not include the vacuum, all of the coefficients with $\sigma_z\neq 0$ will be 0 and the system will converge to the W-state.

\section{Absorbing State Phase Transition}
\label{Sec: Absorbing Phase Transition Appendix}
The matrix elements for ${\cal L}_1$, ${\cal L}_2$, and ${\cal L}_3 + {\cal L}_{3'}$, after renormalizing by ${\cal N}(q,q_z,\sigma_z)$, are given by

\begin{gather}
\begin{split}
    &{\cal L}_{1}^{(N)}(q-1,q_z;q,q_z)
    =  2 \gamma_1 \left[\left(\frac{N}{2} - q + 1 \right)^2 - \sigma_z^2 \right] \\
    &{\cal L}_{1}^{(N)}(q,q_z;q,q_z)
    = -2\gamma_1 \left[ \left(\frac{N}{2} - q\right)^2 -  \sigma_z^2 \right]
\end{split}
\label{eq: L1 renormalized elements}
\end{gather}

\begin{gather}
\begin{split}
    &{\cal L}_{2}^{(N)}(q,q_z-1;q,q_z)
    =  \gamma_2 (q+q_z-1)(q-q_z+1) \\
    &{\cal L}_{2}^{(N)}(q-1,q_z-1;q,q_z)
    =  \gamma_2 \left[\left(\frac{N}{2} - q + 1 \right)^2 - \sigma_z^2 \right] \\
    &{\cal L}_{2}^{(N)}(q,q_z;q,q_z)
    = -\gamma_2 \left(\frac{N}{2} + q_z \right) \left(\frac{N}{2} + q_z -1\right) - \gamma_2 \sigma_z^2
\end{split}
\label{eq: L2 renormalized elements}
\end{gather}

\begin{gather}
\begin{split}
    &{\cal L}_{3}^{(N)}(q,q_z+1;q,q_z)
    =  \gamma_3 (q+q_z)(q+q_z+1) \\
    &{\cal L}_{3}^{(N)}(q,q_z-1;q,q_z)
    =  \gamma_3 (q+q_z-1)(q-q_z+1) \\
    &{\cal L}_{3}^{(N)}(q+1,q_z+1;q,q_z)
    =  -\gamma_3 (q+q_z + 2)(q+q_z+1)\\
    &{\cal L}_{3}^{(N)}(q-1,q_z-1;q,q_z)
    =  -\gamma_3 \left[\left(\frac{N}{2} - q + 1 \right)^2 - \sigma_z^2 \right] \\
    &{\cal L}_{3}^{(N)}(q+1,q_z;q,q_z)
    =  \gamma_3 (q+q_z+1)(q-q_z+1) \\
    &{\cal L}_{3}^{(N)}(q-1,q_z;q,q_z)
    =  \gamma_3 \left[\left(\frac{N}{2} - q + 1 \right)^2 - \sigma_z^2 \right]\\
    &{\cal L}_{3}^{(N)}(q,q_z;q,q_z)
    = -\gamma_3 \left[(N-1)\left(\frac{N}{2} + q_z \right) - 2q \left(\frac{N}{2}-q \right) \right]
\end{split}
\label{eq: L3 renormalized elements}
\end{gather}

It is straightforward to check from these matrix elements that the W-state, written in terms of renormalized generalized Dicke states as 

\begin{gather}
    {\cal D}_W^{(N)} = \frac{1}{N}{|\frac{N}{2} - 1,-\frac{N}{2} + 1,0\rangle}^{(N)} + \frac{1}{N}{|\frac{N}{2},-\frac{N}{2} + 1,0\rangle}^{(N)},
\end{gather}
is individually a steady state of ${\cal L}_{1}$, ${\cal L}_{2}$, and ${\cal L}_{3}$.  Furthermore, the vacuum $|\frac{N}{2},-\frac{N}{2},0 \rangle$ is also a steady state. 
The Lindbladian ${\cal L}_3$, on the other hand, has another steady state - the maximally mixed state (within the subspace of all states excluding the vacuum).  Writing in the renormalized generalized Dicke basis, the maximally mixed (sans vacuum) state is given by

\begin{gather}
    {\cal D}_{\text{mixed}}^{(N)} = \sum_{q_z\neq -\frac{N}{2}} \frac{1}{2^N - 1} {|\frac{N}{2},q_z,0\rangle}^{(N)}.
    \label{eq: Dicke Mixed}
\end{gather}
Using \eqref{eq: L3 renormalized elements}, we have ${\cal L}_3^{(N)} {\cal D}_{\text{mixed}}^{(N)} = 0$, i.e. it is indeed a steady state. 

We remark that a phase transition will still occur even if $\gamma_2 = 0$.  This may be understood in the following way.  If $\gamma_1 \gg \gamma_3$, then between each quantum jump from $\gamma_3$ terms, the contribution from $\gamma_1$ will push each $q_z$ sector to the corresponding pure Dicke state.  If the pure Dicke state is reached for a given $q_z$ sector, then ${\cal L}_{3'}$ can no longer raise $q_z$ (as Dicke states contain no singlets) and ${\cal L}_3$ lower $q_z$ out of that sector.  Thus, now $\gamma_3$ acts to lower $q_z$ until the system is within the 1 particle subspace.  The addition of the ${\cal L}_{2}$ term then only serves to increase the critical value of $\gamma_3$ where the transition takes place.  

To investigate the phase transition, the Lindbladian---from the matrix elements \eqref{eq: L1 renormalized elements}, \eqref{eq: L2 renormalized elements}, and \eqref{eq: L3 renormalized elements}---is exactly diagonalized in Fig. \ref{fig: DPT Numerics} of the main manuscript as well as used to evolve the system starting initially in a fully polarized spin-up state.  

To analyze the critical behavior of the phase transition, it is helpful to numerically investigate how the expected magnetization density evolves with time beginning from the fully polarized spin-up state. 
 Namely, we analyze the expectation value $\langle M(t)\rangle = \Tr{M \rho(t)}$ where

 \begin{gather}
     M = \frac{1}{N} \sum_a \frac{1 - \sigma^z_a}{2}. 
 \end{gather}

\begin{figure}
    \centering
    \includegraphics[width=0.6\textwidth]{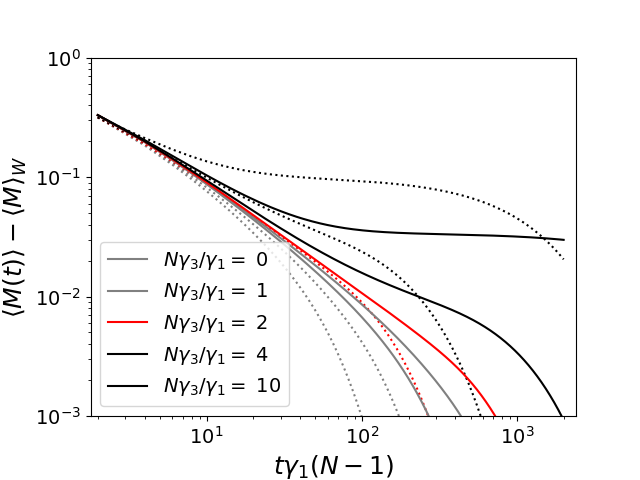}
    \caption{Decay of magnetization density near the critical point.  Dashed lines represent numerics with $N=50$ sites and solid lines $N=200$. 
  Lindblad jump rates are fixed such that $\gamma_1 = \gamma_2$, $\gamma_3 = \gamma_{3'}$.  Critical behavior appears at $N \gamma_3 / \gamma_1 \approx 2$ where the system exhibits polynomial decay.  For smaller $N \gamma_3/\gamma_1$, the system decays exponentially to the W-state, while for larger $N \gamma_3/\gamma_1$ the system approaches the mixed phase---corresponding to a constant magnetization density larger than $\langle M \rangle_W$.  Finite size effects lead to the eventual exponential decay into the W-state at long times for any finite $N \gamma_3/\gamma_1$.}  
    \label{fig: Magnetization Density}
\end{figure}

In Fig. \ref{fig: Magnetization Density} we plot the magnetization density minus the magnetization density of the W-state $\langle M \rangle_W = \Tr{|W\rangle \langle W| M} = \frac{1}{N}$.  We see a clear separation between dynamics for weak, error rates $\gamma_3$ where the magnetization density exponentially decays to that of the W-state, and dynamics for strong $\gamma_3$ where the magnetization approaches a constant value larger than the W-state magnetization density (corresponding to that of the mixed phase).  Polynomial (linear on the log-log plot) decay at $N \gamma_3/\gamma_1 \approx 2$ provides an estimate of the critical point.  The slope corresponds to the decay exponent and is given by $\delta=0.93$.  This is roughly similar to the mean-field directed percolation critical exponent $\delta=1$, which---due to the all-to-all nature of the model---would be consistent with the preparation procedure being of the DP type.  One might expect, for system sizes on the order of 200 sites, that the estimate for $\delta$ should be more precise, but this is not the case for a few reasons outlined below.  It is then difficult to conclude, at this point, if the value $\delta=0.93$ is truly distinct from $1$ or simply due to the numerical limitations of the small system size.   

The subtleties regarding figure \ref{fig: Magnetization Density} are as follows.  First, in a finite system, the dynamics will always converge to the W-state in the long time limit---even for values of $N \gamma_3/\gamma_1$ that are deep within the mixed phase in the thermodynamic limit.  To see why this is the case, consider the maximally mixed state.  The expectation value to be in the 1 spin-up or 2 spin-up sectors is exponentially small in system size---namely, given by the number of states in that sector divided by the total number of states, i.e. $\frac{\binom{N}{1} - 1}{2^N - 2}$ and $\frac{\binom{N}{2}}{2^N - 2}$ respectively.  For finite systems, these expectation values are non-zero and quantum jumps $L_1$ and $L_2$ will take states from these subspaces into the (dark) W-state.  This means even for values of $N \gamma_3/\gamma_1$ deep within the mixed state phase, finite system numerics exhibit exponential decay into the magnetization density of the W-state at long times (as seen in Fig. \ref{fig: Magnetization Density}).  It is the dynamics before this exponential decay takes over that provides insight into the expected behavior in the thermodynamic limit.

This problem becomes especially relevant near the critical point.  This is because the Poisson distribution of the steady state gets shifted to lower and lower $S_z$.  Thus the finite size, exponential decay to the W-state becomes increasingly prominant for $\gamma_3$ approaching the critical point.  Thus---even at system sizes of $200$ sites---distinguishing between exponential decay to the W-state due to this finite size effect and exponential decay to the W-state due to truly being in the W-state preparation phase can make it difficult to accurately pin down the critical point or provide accurate estimates for critical exponents.

We also note that, for this reason, Fig. \ref{fig: DPT Numerics} in the main letter likely provides an overestimate of the location of the critical point.  Near the critical point, the steady-state magnetization density of the mixed phase is low---implying the exponential decay to the W-state will happen much faster for finite system sizes.  This is why values near the critical point, such as $N \gamma_3/\gamma_1 = 10$, still appear to be in the ordered phase in Fig. \ref{fig: DPT Numerics} despite \ref{fig: Magnetization Density} suggesting it should be in the mixed phase---the exponential decay (due to finite-size) to the W-state has already become relevant on the numerical time scales.    

Another comment on Fig. \ref{fig: Magnetization Density}: the rate at which quantum jumps act on a single spin changes as system size is increased in all-to-all systems.  For example, the rate at which a spin experiences an $L_1$ quantum jump is $(N-1) \gamma_1$.  Therefore, we plot $t (N-1) \gamma_1$ to properly compare dynamics for different system sizes in Fig. \ref{fig: Magnetization Density}.

\section{Coherence in the Mixed Steady State}
We derive the expectation value $\langle X_a X_b \rangle$ in terms of generalized Dicke states for all-to-all systems.  First, we use permutation symmetry to rewrite $\langle X_a X_b \rangle$ in terms of all-to-all operators.  Namely,

\begin{gather}
    \langle X_a X_b \rangle = \frac{1}{N(N-1)} \sum_{a\neq b} \langle X_a X_b \rangle = \frac{1}{N(N-1)} \left[\langle \left(\sum_a X_a \right)  \left(\sum_b X_b \right)\rangle - N  \right].
    \label{eq: XX start}
\end{gather}

Let us now rewrite $\langle \left(\sum_a X_a \right)  \left(\sum_b X_b \right)\rangle$ in terms of generalized Dicke states and operators.  Note that 

\begin{gather}
    \left( \sum_a X_a \right) \rho = \left({\cal M}_+ + {\cal N}_+ + {\cal M}_- + {\cal N}_-   \right) [\rho]
\end{gather}
and that $\Tr = \delta_{q,0} \delta_{\sigma_z,0} \sum_{q_z} \langle q,q_z,\sigma_z|$.  We therefore have

\begin{gather}
    \langle \left(\sum_a X_a \right)  \left(\sum_b X_b \right)\rangle = \delta_{q,0} \delta_{\sigma_z,0} \sum_{q_z} \alpha_{q,q_z,\sigma_z} \langle q,q_z,\sigma_z| \left({\cal M}_+ + {\cal N}_+ + {\cal M}_- + {\cal N}_-   \right)^2 | q,q_z,\sigma_z \rangle  \\
    = N + 2 \sum_{q_z} \sum_{\ell=-1}^1 \alpha_{\frac{N}{2}-1,q_z+\ell,\ell}.
    \label{eq: XX permutations}
\end{gather}

Therefore, combining equations \eqref{eq: XX start} and \eqref{eq: XX permutations} we have

\begin{gather}
    \langle X_a X_b \rangle = \frac{1}{\binom{N}{2}} \sum_{q_z} \sum_{\ell = -1}^1 \alpha_{\frac{N}{2}-1,q_z+\ell,\ell}.
\end{gather}


\section{Nearest-Neighbor Results}
\subsection{Transition at $O(\frac{1}{Z}$)}
We here elaborate on why the transition occurs when $\frac{\gamma_3}{\gamma_1} = O(\frac{1}{Z})$ for a lattice with uniform coordination number $Z$.  As discussed in the main letter, it is helpful to consider the expectation value $\langle P^{|1,0\rangle,|1,-1 \rangle} \rangle$, where

\begin{gather}
    P^{|1,0\rangle,|1,-1 \rangle} = \frac{1}{N-1} \sum_{\langle a,b \rangle} P^{|1,0\rangle,|1,-1 \rangle}_{a,b} \\
    \frac{1}{N-1} \sum_{\langle a,b \rangle} \left( |1,0\rangle_{a,b} \langle 1,0|_{a,b} + |1,-1\rangle_{a,b} \langle 1,-1|_{a,b} \right) \\
    = \frac{1}{N-1} \sum_{\langle a,b \rangle} \frac{1}{2} \left( P_a^0 + P_b^0 + \sigma_a^+ \sigma_b^- + \sigma_a^- \sigma_b^+ \right) \label{eq: P_TO}
\end{gather}
where $P_a^0 = \left(\frac{1 + Z_a}{2} \right)$ is the projector of site $a$ onto the state $\downarrow$.  We similarly define $P^{|S\rangle} = \frac{1}{N-1} \sum_{\langle a,b \rangle} P^{|S\rangle}_{a,b}$ with $P^{|S\rangle}_{a,b} = |S\rangle_{a,b} \langle S|_{a,b}$ for any two-site state $|S\rangle \in \left\{|0,0\rangle, |1,-1\rangle, |1,0\rangle, |1,1\rangle \right\}$.  

The evolution of $\langle P^{|1,0\rangle,|1,-1 \rangle} \rangle$ is given by

\begin{gather}
    \frac{d}{dt} \langle P^{|1,0\rangle,|1,-1 \rangle} \rangle = \Tr \left\{ P^{|1,0\rangle,|1,-1 \rangle} \frac{d \rho}{dt} \right\} \\
    = \frac{1}{N-1}  \sum_{\substack{\langle a,b \rangle\\ j \in \left\{1,2,3,3' \right\}}} \sum_{\langle c,d \rangle} \gamma_j \Tr \left\{\rho \left[ L_{j;a,b}^\dagger P_{c,d}^{|1,0\rangle,|1,-1 \rangle} L_{j;a,b} - \frac{1}{2} \left(L_{j;a,b}^\dagger L_{j;a,b} P_{c,d}^{|1,0\rangle,|1,-1 \rangle} + P_{c,d}^{|1,0\rangle,|1,-1 \rangle} L_{j;a,b}^\dagger L_{j;a,b} \right)  \right] \right\} \label{eq: evolution P expectation}
\end{gather}

Let us first discuss the case where $a=c$ and $b=d$.  For cohering terms like $L_1$, we have that

\begin{gather}
    L_{1;a,b}^\dagger P^{|1,0\rangle,|1,-1 \rangle}_{a,b} L_{1;a,b} - \frac{1}{2} \left(L_{1;a,b}^\dagger L_{1;a,b} P^{|1,0\rangle,|1,-1 \rangle}_{a,b} + P^{|1,0\rangle,|1,-1 \rangle}_{a,b} L_{1;a,b}^\dagger L_{1;a,b} \right) \\
    = |0,0\rangle \langle 1,0| P^{|1,0\rangle,|1,-1 \rangle} |1,0\rangle \langle 0,0| - \frac{1}{2} \left(|0,0\rangle \langle 0,0| P^{|1,0\rangle,|1,-1 \rangle} + P^{|1,0\rangle,|1,-1 \rangle} |0,0\rangle \langle 0,0| \right) \\
    = |0,0\rangle \langle 0,0|
\end{gather}
where in the second line we dropped the $a,b$ subscripts for brevity.  Thus, the contribution from the $L_1$ on $a,b$ and $P^{|1,0\rangle,|1,-1 \rangle}_{a,b}$ term in the sum \eqref{eq: evolution P expectation} is

\begin{gather}
    \gamma_1 \Tr \left\{\rho |0,0\rangle \langle 0,0| \right\} \\
    = \gamma_1 \langle P^{|0,0\rangle}_{a,b} \rangle.
\end{gather}
Similarly, the contribution from $L_2$ on $a,b$ is $\gamma_2 \langle P^{|1,1\rangle}_{a,b} \rangle$.  For the decohering terms (for example, taking first $L_{3'}$ and again dropping $a,b$ subscripts for brevity), however, we have that

\begin{gather}
    L_{3'}^\dagger P^{|1,0\rangle,|1,-1 \rangle} L_{3'} - \frac{1}{2} \left(L_{3'}^\dagger L_{3'} P^{|1,0\rangle,|1,-1 \rangle} + P^{|1,0\rangle,|1,-1 \rangle} L_{3'}^\dagger L_{3'} \right) \\
    = |0,0\rangle \langle 1,1| P^{|1,0\rangle,|1,-1 \rangle} |1,1\rangle \langle 0,0| - \frac{1}{2} \left(|0,0\rangle \langle 0,0| P^{|1,0\rangle,|1,-1 \rangle} + P^{|1,0\rangle,|1,-1 \rangle} |0,0\rangle \langle 0,0| \right) \\
    = 0.
\end{gather}
The contribution from $L_{3'}$ is similarly $0$.  We therefore have that the contribution to \eqref{eq: evolution P expectation} from the $a=b$,$c=d$ terms (after summing over $a,b$), denoted by $\frac{d}{dt} \langle P^{|1,0\rangle,|1,-1 \rangle} \rangle |_{a=b,c=d}$, is given by 

\begin{gather}
    \frac{d}{dt} \langle P^{|1,0\rangle,|1,-1 \rangle} \rangle |_{a=b,c=d} = \gamma_1 \langle P^{|0,0\rangle} \rangle + \gamma_2 \langle P^{|1,1\rangle} \rangle
\end{gather}

We now consider when it is not the case that $a=c$ and $b=d$.  If $a\neq c,d$ and $b\neq c,d$, then $[L_{\alpha;a,b},P^{|1,0\rangle,|1,-1 \rangle}_{c,d}] = 0$, implying these terms in the sum \eqref{eq: evolution P expectation} are 0.  Now consider terms where either $a=c$ or $b=d$.  Specifically, for a jump operator acting on the sites $a,b$, we must consider the projectors $\sum_{\substack{d \\ s.t. \langle a,d \rangle}}P^{|1,0\rangle,|1,-1 \rangle}_{a,d} + \sum_{\substack{c \\ s.t. \langle b,c \rangle}} P^{|1,0\rangle,|1,-1 \rangle}_{b,c}$.  From now on we will write $\sum_c$ or $\sum_d$ for brevity.  Before we calculate the contribution to the sum \eqref{eq: evolution P expectation} from these terms, it is helpful to establish a few identities.  Namely,

\begin{subequations}
\begin{gather}
\begin{split}
    &\left\{ |0,0\rangle_{a,b} \langle 0,0|_{a,b} ,\left(\sum_d P^{|1,0\rangle,|1,-1 \rangle}_{a,d} + \sum_c P^{|1,0\rangle,|1,-1 \rangle}_{b,c}\right) \right\}  \\
    = &\left( Z-1 + \sum_c P_c^0  + \sum_d P_d^0 \right) |0,0\rangle_{a,b} \langle 0,0|_{a,b} \\ 
    + &\frac{1}{2 \sqrt{2}} \left[|1,0\rangle_{a,b} \langle 0,0|_{a,b} (\sum_c \sigma_c^- - \sum_d \sigma_d^-) - |1,0\rangle_{a,b} \langle 1,1|_{a,b} (\sum_c \sigma_c^+ - \sum_d \sigma_d^+) + h.c.  \right]
\end{split}
\end{gather}
\begin{gather}
\begin{split}
    &\left\{ |1,1\rangle_{a,b} \langle 1,1|_{a,b} , \left(\sum_d P^{|1,0\rangle,|1,-1 \rangle}_{a,d} + \sum_c P^{|1,0\rangle,|1,-1 \rangle}_{b,c}\right) \right\} \\
    = \left(\sum_c P_c^0 + \sum_d P_d^0 \right) &|1,1\rangle_{a,b} \langle 1,1|_{a,b} + \frac{1}{2} \left[|1,1\rangle_{a,b} \left( \langle 10|_{a,b} \sum_c \sigma_c^- + \langle 01|_{a,b} \sum_d \sigma_d^-\right) + h.c. \right]
\end{split}
\end{gather}
\begin{gather}
    \langle 1,0|_{a,b} \left(\sum_d P^{|1,0\rangle,|1,-1 \rangle}_{a,d} + \sum_c P^{|1,0\rangle,|1,-1 \rangle}_{b,c}\right) |1,0 \rangle_{a,b} = \frac{1}{2} \left(Z-1 + \sum_c P_c^0 + \sum_d P_d^0 \right),
\end{gather}
\begin{gather}
    \langle 0,0|_{a,b} \left(\sum_d P^{|1,0\rangle,|1,-1 \rangle}_{a,d} + \sum_c P^{|1,0\rangle,|1,-1 \rangle}_{b,c}\right) |0,0 \rangle_{a,b} =  \frac{1}{2} \left(Z-1 + \sum_c P_c^0 + \sum_d P_d^0 \right),
\end{gather}
\begin{gather}
    \langle 1,1|_{a,b} \left(\sum_d P^{|1,0\rangle,|1,-1 \rangle}_{a,d} + \sum_c P^{|1,0\rangle,|1,-1 \rangle}_{b,c}\right) |1,1 \rangle_{a,b} =  \frac{1}{2} \left(\sum_c P_c^0 + \sum_d P_d^0 \right),
\end{gather}
\label{eq: P identities}
\end{subequations}

where in the above relations we have used that $P^{|1,0\rangle,|1,-1 \rangle}_{a,b} = \frac{1}{2} \left( P_a^0 + P_b^0 + \sigma_a^+ \sigma_b^- + \sigma_a^- \sigma_b^+ \right)$.

Using \eqref{eq: P identities}, we find the following for $L_1$, $L_2$, $L_3$, and $L_{3'}$:

\begin{center}
\begin{tabular}{c|c}
     $L$ &  $L_{a,b}^\dagger \left(\sum_d P^{|1,0\rangle,|1,-1 \rangle}_{a,d} + \sum_c P^{|1,0\rangle,|1,-1 \rangle}_{b,c}\right) L_{a,b} - \frac{1}{2} \left\{L_{a,b}^\dagger L_{a,b}, \sum_d P^{|1,0\rangle,|1,-1 \rangle}_{a,d} + \sum_c P^{|1,0\rangle,|1,-1 \rangle}_{b,c} \right\}$\\
     \hline
     $L_1$ & $-\frac{1}{4 \sqrt{2}} \left[|1,0\rangle_{a,b} \langle 0,0|_{a,b} (\sum_c \sigma_c^- - \sum_d \sigma_d^-) - |1,0\rangle_{a,b} \langle 1,1|_{a,b} (\sum_c \sigma_c^+ - \sum_d \sigma_d^+) + h.c.  \right]$ \\
     $L_2$ & $\frac{Z-1}{2} |1,1\rangle_{a,b} \langle 1,1|_{a,b} - \frac{1}{4} \left[|1,1\rangle_{a,b} \left( \langle 10|_{a,b} \sum_c \sigma_c^- + \langle 01|_{a,b} \sum_d \sigma_d^-\right) + h.c. \right]$\\
     $L_3$ & $\frac{Z-1}{2} |1,1\rangle_{a,b} \langle 1,1|_{a,b} - \frac{1}{4} \left[|1,1\rangle_{a,b} \left( \langle 10|_{a,b} \sum_c \sigma_c^- + \langle 01|_{a,b} \sum_d \sigma_d^-\right) + h.c. \right]$\\
     $L_{3'}$ & $- \frac{Z-1}{2} |0,0\rangle_{a,b} \langle 0,0|_{a,b} -\frac{1}{4 \sqrt{2}} \left[|1,0\rangle_{a,b} \langle 0,0|_{a,b} (\sum_c \sigma_c^- - \sum_d \sigma_d^-) - |1,0\rangle_{a,b} \langle 1,1|_{a,b} (\sum_c \sigma_c^+ - \sum_d \sigma_d^+) + h.c.  \right]$\\
\end{tabular}
\end{center}

In the above table, the term $\sum_c \sigma_c^- - \sum_d \sigma_d^-$ and its hermitian conjugate appear several times.  In the all-to-all limit, we have that $\sum_c \sigma_c^- - \sum_d \sigma_d^- = 0$ as the neighbors of site $a$ and site $b$ are the same (i.e. all sites).  We now explain why these terms are also negligible for the NN model in the thermodynamic limit.  Let us first look at the term $|1,0\rangle_{a,b} \langle 0,0|_{a,b} (\sum_c \sigma_c^- - \sum_d \sigma_d^-)$.  This term's contribution in equation \eqref{eq: evolution P expectation} is $\frac{1}{N-1}\sum_{\langle a,b \rangle} \Tr \left\{ \langle 0,0|_{a,b} \rho |1,0\rangle_{a,b} (\sum_c \sigma_c^- - \sum_d \sigma_d^-)  \right\}$.  For a general, random state, this is $0$ as the state will on average have no asymmetry between the neighbors $d$ of $a$ and the neighbors $c$ of $b$.  Specifically, suppose $\Tr \left\{ \langle 0,0|_{a,b} \rho |1,0\rangle_{a,b} \sum_c \sigma_c^- \right\} = \xi_{abc}$ where $\xi_{abc}$ is some normal distribution with mean $\mu_{abc}$ and variance $\Var(\xi_{abc})$ determined by the random state $\rho$ (and similarly $\xi_{abd}$ for the $\sigma_d^-$ expectation value).  We therefore have $\Tr \left\{ \langle 0,0|_{a,b} \rho |1,0\rangle_{a,b} (\sum_c \sigma_c^- - \sum_d \sigma_d^-)  \right\} = \xi_{abc} - \xi_{abd} \equiv \xi_{ab}$ where $\xi_{ab}$ is a normal distribution with $\mu_{ab} = \mu_{abc} - \mu_{abd}$ and $\Var(\xi_{ab}) = \Var(\xi_{abc}) + \Var(\xi_{abd})$.  If $\xi_{abc}$ and $\xi_{abd}$ are i.i.d. we have that $\mu_{ab} = 0$ and $\Var(\xi_{ab}) = 2 \Var(\xi_{abc})$.  Summing over $a,b$ we get $\frac{1}{N-1}\sum_{\langle a,b \rangle} \Tr \left\{ \langle 0,0|_{a,b} \rho |1,0\rangle_{a,b} (\sum_c \sigma_c^- - \sum_d \sigma_d^-)  \right\} = \frac{1}{N-1} \sum_{\langle a,b \rangle} \xi_{ab} \equiv \xi$ where (again assuming i.i.d. of $\xi_{ab}$) $\xi$ has mean $\mu=0$ and variance $\Var(\xi) = \frac{2}{N-1} \Var(\xi_{abc})$.  Thus, $\xi$ is precisely $0$ in the thermodynamic limit.  Therefore, we will treat the terms $\sum_c \sigma_c^- - \sum_d \sigma_d^-$ as negligible.     

Now equipped with the solution to $ L_{a,b}^\dagger P^{|1,0\rangle,|1,-1 \rangle}_{c,d} L_{a,b} - \frac{1}{2} \left(L_{a,b}^\dagger L_{a,b} P^{|1,0\rangle,|1,-1 \rangle}_{c,d} + P^{|1,0\rangle,|1,-1 \rangle}_{c,d} L_{a,b}^\dagger L_{a,b} \right) $ for all possible values of $a$, $b$, $c$, and $d$, we find that \eqref{eq: evolution P expectation} becomes

\begin{gather}
    \frac{d}{dt} \langle P^{|1,0\rangle,|1,-1 \rangle} \rangle = \gamma_1 \langle P^{|0,0\rangle} \rangle + \gamma_2 \langle P^{|1,1\rangle} \rangle - \gamma_{3'} \frac{Z-1}{2} \langle P^{|0,0\rangle} \rangle + \left(\gamma_2 + \gamma_3 \right) \frac{Z-1}{2} \langle P^{|1,1\rangle} \rangle - \frac{\gamma_2 + \gamma_3}{2} \langle T \rangle
    \label{eq: final expect P}
\end{gather}
where $T = \frac{1}{N-1} \sum_{\langle a,b \rangle} \frac{1}{4} \left[|1,1\rangle_{a,b} \left( \langle 10|_{a,b} \sum_c \sigma_c^- + \langle 01|_{a,b} \sum_d \sigma_d^-\right) + h.c. \right]$.

The terms $\gamma_1 \langle P^{|0,0\rangle} \rangle + \gamma_2 \langle P^{|1,1\rangle} \rangle$ come from the cohering quantum jumps acting in cases $a=c$, $b=d$ and correspond to an increase in the coherence of the system.  The term $\left(\gamma_2 + \gamma_3 \right) \frac{Z-1}{2} \langle P^{|1,1\rangle} \rangle$ corresponds to the $\gamma_2$ and $\gamma_3$ terms lowering total $S_z$ and thereby increasing the number of pairs in the $|1,-1\rangle$ state.  The rest of the terms correspond to cases where a coherent state on any two sites $a,b$ is destroyed by jump operators which act between either $a$ or $b$ and a neighboring site. 

For an initial density matrix in the single spin-up sector, equation \eqref{eq: final expect P} becomes \eqref{eq: ddt of P} in the main text.

\subsection{Dynamical Exponents and Exact Results for Lindblad Spectrum}
Solving the Lindbladian for the 1D system is, in general, hard.  However, as we will show in this section, it is possible to solve explicitly for several important classes of eigenstates and eigenvalues.  This allows us to show explicitly that preparation time of the W-state is $O(N^2)$, namely we show the dynamical exponent $\Delta \sim N^{-z}$ deep within the W-state phase is diffusive: $z=2$.  We first, however, find a class of eigenstates which exist for general $\gamma_3$.  

\subsubsection{Exact Solution for Eigenmatrices with Off-diagonal Correlations to Dark States}
\label{Sec: exact solution to off diag correlations with dark}
To find a first class of exact eigendensity-matrices, we re-write the Lindbladian as

\begin{gather}
    {\cal L}[\rho] = \sum_\ell \gamma_\ell L_\ell \rho L_\ell^\dagger + \left\{{\cal H}_{XX},\rho \right\} 
\end{gather}
where (for $\gamma_1=\gamma_2$ and $\gamma_3 = \gamma_{3'}$) we have
\begin{gather}
    {\cal H}_{XX} = - \frac{\gamma_1 + \gamma_3}{2} \sum_{\langle a,b \rangle} |1,1\rangle_{a,b}\langle 1,1|_{a,b} + |0,0\rangle_{a,b}\langle 0,0|_{a,b} \\
    = - \frac{\gamma_1 + \gamma_3}{2} \left[ \sum_a \frac{1-Z_a}{2} - \frac{1}{2} \sum_{\langle a,b \rangle} \left( \sigma_a^+ \sigma_b^- + \sigma_a^- \sigma_b^+ \right)  \right].
\end{gather}

Letting ${\cal H}_{XX} |\lambda_\alpha \rangle = \lambda_\alpha | \lambda_\alpha \rangle$ be the eigenstates and eigenvalues for ${\cal H}_{XX}$, we have that $|\lambda_\alpha\rangle \langle 0|$, $|\lambda_\alpha\rangle \langle W|$, and their hermitian conjugates must all be eigenmatrices for ${\cal L}$ with corresponding eigenvalue $\lambda_\alpha$ (we will refer to the space of density matrices spanned by $|\lambda_\alpha\rangle \langle 0|$, $|\lambda_\alpha\rangle \langle W|$, $|0\rangle \langle \lambda_\alpha|$, and $|W\rangle \langle \lambda_\alpha|$ as ${\cal L}_{\cal H}$).  This is because $|0\rangle$ and $|W\rangle$ are dark states, thus $L|W\rangle = {\cal H}_{XX} |W\rangle = 0$ and ${\cal L}[|\lambda_\alpha \rangle \langle W |] = \lambda_\alpha |W\rangle \langle \lambda_\alpha |$ (similarly for $|\lambda_\alpha \rangle \langle 0|$ and hermitian conjugates).  Thus, the solution to ${\cal H}_{XX}$ provides a class of solutions to the Lindbladian.

Here, we note ${\cal H}_{XX}$ is the XX-model and may be exactly solved on the 1D chain using a Jordan-Wigner transformation followed by a Fourier transform \cite{Lieb1961XX}.  In particular, we have that after diagonalizing

\begin{gather}
    {\cal H}_{XX} = \sum_k E_k c_k^\dagger c_k, \;\; E_k = - \frac{\gamma_1 + \gamma_3}{2} \left[1 - \cos (k) \right]
\end{gather}       
with---for integers $m=0,1,...,N-1$ and total number of jordan-wigner fermions $N_f$ in the system---
\begin{gather}
    k = \begin{cases}
        \frac{2 \pi}{N} m & N_f \text{ Odd }\\
        \frac{2 \pi}{N} \left(m + \frac{1}{2} \right) & N_f \text{ Even }
    \end{cases}.
\end{gather}

The lowest energy excitations (which will be relevant for the gap of the Lindbladian) occur when k is close to $0$ (mod $2 \pi$).  Expanding in this regime, 
\begin{gather}
    E_k  \stackrel{\text{small } k}{\rightarrow} - \frac{\gamma_1 + \gamma_3}{4} k^2 \propto N^{-2}.
\end{gather}
Within the subspace ${\cal L}_{\cal H}$, the gap for the Lindbladian is $\Delta \sim N^{-2}$ and thus the dynamical exponent is given by $2$.  The gap within the subspace ${\cal L}_{\cal H}$---given by $\Delta = (\gamma_1 + \gamma_3) \left[1 - \cos(\frac{\pi}{N}) \right]$ and corresponding to the density matrices with $|\lambda_\alpha \rangle = c_{\pi/N}^\dagger c_{2\pi - \pi/N}^\dagger |0\rangle$---also matches the gap found by exactly diagonalizing the full Lindbladian for numerically accessible system sizes (and for parameters far away from the critical point).  This naively suggests the dynamical exponent far from the critical point is $2$, however it is important to check whether the exact solutions to ${\cal L}$ found play a role in the dynamics of the system for physically relevant initial states.  For example, starting from the fully polarized spin-up state, this is certainly not the case deep within the mixed phase as the dynamics of the system will never approach the W-state or the vacuum, and so the dynamics occurs completely in the space of density matrices orthogonal to ${\cal L}_{\cal H}$. 

In fact, for the initial fully polarized spin up state, many of the exact eigenmatrix solutions found do not play a role deep within the ordered, W-state preparation phase either.  This is because no jump operators in the Lindbladian ${\cal L}$ generate correlations between different total magnetization sectors.  Thus, all matrix elements between states with different total magnetization will be zero in the density matrix when starting from the fully polarized spin-up state.  Thus, in this case, the only relevant eigenmatrices in the space ${\cal L}_{\cal H}$ come from the single fermion solutions $|\lambda_\alpha \rangle$.  In this case, the largest non-zero eigenvalue is $-\frac{\gamma_1 + \gamma_3}{2} \left[1 - \cos(\frac{2 \pi}{N}) \right]$ and corresponds to the eigenmatrix $|\lambda_\alpha \rangle \langle W|$ (and hermitian conjugate) with $|\lambda_\alpha \rangle = c_{2 \pi/N}^\dagger |0\rangle$.  However, this gap $\frac{\gamma_1 + \gamma_3}{2} \left[1 - \cos(\frac{2 \pi}{N}) \right]$ does not match the gap found numerically by starting from the fully polarized spin-up state and looking at the exponential decay of magnetization at long times.  In other words, there is another eigenmatrix which is relevant to the dynamics, lies outside the space ${\cal L}_{\cal H}$, and with a corresponding eigenvalue given by that found numerically from the long time magnetization decay.     

\subsubsection{Single Spin-up Solutions Deep in Ordered Phase}
\label{sec: single spin solution}
We now find the largest non-zero eigenvalue relevant to magnetization decay from the fully polarized spin-up state deep within the ordered, W-state preparation phase.  Namely, we are interested in the limit $\gamma_3 \rightarrow 0$.  This corresponds to the preparation time of the W-state in the limit of $0$ errors. 
 Here, since we know the W-state is being prepared, we expect the slowest decaying eigenmatrices to be contained within the single spin-up sector.  This is because the only jump operators acting are $L_1$ and $L_2$, which means the slowest decaying modes should be in either the two spin-up or one spin-up sectors.  The jumps $L_2$ need only find two spin-ups on neighboring sites to lower into the one spin-up space, while $L_1$ needs to generate coherence throughout the system.  Therefore, we expect the slowest decaying mode to live within the single spin-up space (note, this argument fails as soon as $\gamma_3 \neq 0$, since $L_{3'}$ terms may take the system back outside of the single spin-up space).  Therefore, all that is required to find the relevant decay rate and gap is to solve the Lindbladian in the single spin-up sector.  We achieve this below by mapping the Lindbladian in the single spin-up space to a 1D impurity model which may be solved exactly using transfer matrix approaches.  From this solution, we find the gap in this space---and therefore the W-state preparation time---is given by
 \begin{gather}
     \Delta \stackrel{\gamma_3 \rightarrow 0}{=} \gamma_1 \left[1 - \cos \left(\frac{\pi}{N-1} \right) \right] = O(N^{-2}).
     \label{eq: single particle gap}
 \end{gather}
Thus, the dynamical exponent deep within the ordered phase is given by $z=2$ and the preparation time for the W-state scales as $N^2$.  

We now solve the $\gamma_3 \rightarrow 0$ Lindblad dynamics  in the single spin-up subspace.  We notate the space of single spin-up states by
\begin{gather}
    |x\rangle \equiv |\downarrow_0 \downarrow_1 ... \downarrow_{x-1} \uparrow_x \downarrow_{x+1} ...\downarrow_{N-1}\rangle.
\end{gather}
Thus, we are looking for density matrices of the form $\rho = \sum_{x,x'} \rho_{xx'}|x\rangle \langle x'|$ that are solutions to \eqref{eq: Lindbladian}.  We will work in the doubled space to vectorize $\rho$, i.e. $\rho \rightarrow \sum_{x,x'} \rho_{xx'} |x\rangle \otimes|x'\rangle$ and we will write $|x\rangle \otimes|x'\rangle \equiv |x\; x'\rangle$ for brevity.  For our model deep within the ordered phase and in the single spin-up sector, the only relevant jump operators are $L_1$, as $L_3$ and $L_{3'}$ may be ignored since $\gamma_3=0$ and $L_2 |x\rangle = 0$.  We note that
\begin{subequations}
\label{eq: L1 in single particle}
\begin{gather}
    L_{1;x,x+1}|x\rangle = |1,0\rangle_{x,x+1} \langle 0,0|_{x,x+1} |x\rangle = -\frac{1}{2} \left(|x\rangle + |x+1\rangle \right), \\
     L_{1;x-1,x}|x\rangle = \frac{1}{2} \left(|x-1\rangle + |x\rangle \right), \\
     L^\dagger_{1;x,x+1} L_{1;x,x+1}|x\rangle = \frac{1}{2} \left(|x\rangle - |x+1\rangle \right), \\
     L^\dagger_{1;x-1,x} L_{1;x-1,x}|x\rangle = \frac{1}{2} \left(|x\rangle - |x-1\rangle \right).
\end{gather}
\end{subequations}

Using \eqref{eq: L1 in single particle}, we find that the full Lindbladian in the double space is given by

\begin{gather}
    {\cal L}_{\text{single spin-up}} = \sum_x L_{1;x,x+1} \otimes L_{1;x,x+1} - \frac{1}{2} \sum_x \left(L^\dagger_{1;x,x+1} L_{1;x,x+1} \otimes I + I \otimes L^\dagger_{1;x,x+1} L_{1;x,x+1} \right) \\
    \begin{aligned}
    = \frac{1}{4} \sum_x &\left\{ \left[2|x\; x\rangle + |x+1\; x\rangle + |x\; x+1\rangle + |x+1\; x+1\rangle + |x-1\; x-1\rangle + |x-1\; x\rangle + |x\; x-1\rangle \right] \langle x\; x|\right. \\
    &\left.- \left[|x\; x\rangle + |x\; x+1\rangle + |x+1\; x\rangle + |x+1\; x+1\rangle \right] \langle x\; x+1|  \right. \\
    &\left. - \left[|x-1\; x-1\rangle + |x-1\; x\rangle + |x\; x-1\rangle + |x\; x\rangle \right]\langle x\; x-1| \right\} \\
    -\frac{1}{2} \sum_{x,x'} &\left(\left[|x\; x'\rangle - \frac{1}{2} |x+1\; x'\rangle -\frac{1}{2} |x-1\; x'\rangle\right] \langle x\; x'| + \left[|x\; x'\rangle - \frac{1}{2} |x\; x'+1\rangle -\frac{1}{2} |x\; x'-1\rangle\right] \langle x\; x'|\right)
    \end{aligned}\\
    \begin{aligned}
    = \frac{1}{4} &\sum_{k_+} \left(\begin{tabular}{c c c}
        $|k_+\; 1\rangle$ & $|k_+\; 0\rangle$ & $|k_+\; -1\rangle$
    \end{tabular} \right) \left(\begin{tabular}{c c c}
         $-1$ & $2 \cos \left(k_+\right)$ & $-1$\\
         $-2 \cos \left(k_+\right)$ & $4\cos^2 \left(k_+\right)$ & $-2 \cos \left(k_+\right)$\\
$-1$ & $2 \cos \left(k_+\right)$ & $-1$
    \end{tabular} \right) \left(\begin{tabular}{c}
        $\langle k_+\; 1|$  \\
         $\langle k_+\; 0|$  \\
         $\langle k_+\; -1|$ 
    \end{tabular} \right) \\
    -&\frac{1}{2} \sum_{k_+,x_-} \left(\begin{tabular}{c c c}
        $|k_+\; x_-+1\rangle$ & $|k_+\; x_-\rangle$ & $|k_+\; x_--1\rangle$
    \end{tabular}\right) \left(\begin{tabular}{c}
          $-\cos \left(k_+\right)$ \\
          $2$ \\
         $-\cos \left(k_+\right)$
    \end{tabular} \right) \langle k_+\; x_-|
    \end{aligned}
    \label{eq: impurity model}
\end{gather}
where in the last step we have applied a transformation to the center of mass frame---i.e. applied the transformation $|x \; x'\rangle \rightarrow |x_+ \; x_-\rangle \equiv |\frac{x+x'}{2} \; \frac{x-x'}{2}\rangle$---and then applied a Fourier transform to the left Hilbert space.  The left Hilbert space is diagonal in Fourier space, and in \eqref{eq: impurity model} the right Hilbert space is written in the form of a single particle 1D impurity model---i.e. the first line in \eqref{eq: impurity model} corresponds to an impurity localized at sites $-1$, $0$, and $1$ and the second line in \eqref{eq: impurity model} corresponds to hopping terms between each `site' $x_-$ and its nearest neighbors.  Thus, to diagonalize ${\cal L}_{\text{single spin-up}}$, we must solve this impurity model.  To do so, we employ a transfer matrix approach.  

The solution to Eq. \eqref{eq: impurity model} is of the form 
\begin{gather}
    {\cal L}_{\text{single spin-up}}\left(|k_+\rangle \otimes \Psi \right) = E \left(|k_+\rangle \otimes \Psi \right)
    \label{eq: single particle eigenproblem}
\end{gather}
where $\Psi \equiv \sum_m \psi_m |m\rangle$.  Combining Eqs. \eqref{eq: impurity model} and \eqref{eq: single particle eigenproblem} we have
\begin{gather}
    \psi_m E = \begin{cases}
        \frac{1}{2} \cos \left(k_+\right) \psi_{m+1} - \frac{5}{4} \psi_m - \frac{1}{4} \psi_{m-2} & m=1\\
        \cos \left(k_+\right) \psi_{m+1} + \left[\cos^2 \left(k_+\right) -1\right]\psi_m + \cos \left(k_+\right) \psi_{m-1} & m=0\\
        -\frac{1}{4} \psi_{m+2} - \frac{5}{4} \psi_m + \frac{1}{2}\cos \left(k_+\right) \psi_{m-1} & m=N-1 \\
        \frac{1}{2} \cos \left(k_+\right) \left[\psi_{m+1} + \psi_{m-1} \right] - \psi_m & \text{Otherwise}
    \end{cases}
    \label{eq: equations for transfer matrices}
\end{gather}

Let the transfer matrices $T_m$ be defined by

\begin{gather}
    \left(\begin{tabular}{c}
         $\psi_{m+1}$ \\
        $\psi_m$
    \end{tabular} \right) = T_m \left(\begin{tabular}{c}
         $\psi_{m}$ \\
        $\psi_{m-1}$
    \end{tabular} \right).
\end{gather}
From Eq. \eqref{eq: equations for transfer matrices} we find
\begin{gather}
\label{eq: transfer matrices}
\begin{aligned}
    T_1 &= \left(\begin{tabular}{c c}
        $B$ & $\frac{A}{B-2A}$ \\
        $1$ & $0$
    \end{tabular} \right) \\
    T_0 &= \left(\begin{tabular}{c c}
        $A$ & $-1$ \\
        $1$ & $0$
    \end{tabular} \right) \\
    T_{N-1} &= \left(\begin{tabular}{c c}
        $\frac{2AB-B^2}{A}$ & $\frac{B-2A}{A}$ \\
        $1$ & $0$
    \end{tabular} \right) \\
    T_{m\neq \{1,0,N-1\}} &= \left(\begin{tabular}{c c}
        $B$ & $-1$ \\
        $1$ & $0$
    \end{tabular} \right)
\end{aligned}
\end{gather}
where $A \equiv \frac{E+1 - \cos^2 \left(k_+\right)}{\cos \left(k_+\right)}$ and $B=\frac{2 (E+1)}{\cos \left(k_+\right)}$.

To solve for the eigenvalues, we note that

\begin{gather}
    \left(\begin{tabular}{c}
         $\psi_{1}$ \\
        $\psi_0$
    \end{tabular} \right) = T \left(\begin{tabular}{c}
         $\psi_{1}$ \\
        $\psi_{0}$
    \end{tabular} \right), \;\;
    T \equiv T_0 T_{N-1} T_{m\neq \{1,0,N-1\}}^{N-3} T_1. 
    \label{eq: total transfer matrix}\\
    \implies \det[T-I] = 0.  
    \label{eq: Det of transfer matrices}
\end{gather}

We remark that
\begin{gather}
\label{eq: Tm is Chebyshev}
    T_{m\neq \{1,0,N-1\}}^{m} = \left(\begin{tabular}{c c}
        $U_m(\frac{B}{2})$ & $-U_{m-1}(\frac{B}{2})$ \\
        $U_{m-1}(\frac{B}{2})$ & $-U_{m-2}(\frac{B}{2})$
    \end{tabular} \right)
\end{gather}
where $U_m(x)$ are Chebyshev polynomials of the second kind.  Combining \cref{eq: transfer matrices,eq: total transfer matrix,eq: Det of transfer matrices,eq: Tm is Chebyshev} (and using the identities $U_m(x)=2xU_{m-1}(x) - U_{m-2}(x)$ and $U_m(x)^2 - U_{m-1}(x)U_{m+1}(x)=1$ for Chebyshev polynomials) we find

\begin{gather}
    1 + U_{N-2}(\frac{B}{2}) + \cos (k_+) U_{N-1}(\frac{B}{2}) = 0 \\
    \implies \sin[\arccos(\frac{E+1}{\cos(k_+)})] + \sin[(N-1)\arccos(\frac{E+1}{\cos(k_+)})] + \cos(k_+) \sin[N\arccos(\frac{E+1}{\cos(k_+)})] = 0.
    \label{eq: transfer matrix final solution}
\end{gather}
where we have used the identity $U_m(\cos \theta)=\frac{\sin((m+1)\theta)}{\sin(\theta)}$.

In summary, the eigenvalues to the dissipative W-state preparation procedure without any errors within the single spin-up space are given by \eqref{eq: transfer matrix final solution} for $k_+ = \frac{2 \pi a}{N}$ with $a \in \{0,1,...,N-1 \}$.  As discussed at the start of this subsection, the preparation time for the W-state is therefore given by the largest non-zero eigenvalue from this solution.  We will not attempt to solve explicitly for all the eigenvalues given by \eqref{eq: transfer matrix final solution}.  Instead, we will focus on finding the most important (i.e. largest) eigenvalues from \eqref{eq: transfer matrix final solution}.  First note that $E=-(1 - \cos(k_+))$ is a solution.  These solutions correspond to our solutions in section \ref{Sec: exact solution to off diag correlations with dark} restricted to the single spin-up sector.  However, we also note that we can find all the solutions for when $k_+ = 0$.  To do so, we make an ansatz that solutions are of the form

\begin{gather}
\label{eq: k_+=0 solution}
    E = \begin{cases}
        -(1-\cos(\frac{a\pi}{N-1})) & \text{if $a$ Odd}\\
        -(1-\cos(\frac{a\pi}{N})) & \text{if $a$ Even}
    \end{cases}
\end{gather}
for $a\in \{0,1,...,N-1\}$.  It is straightforward to verify that indeed \eqref{eq: k_+=0 solution} are solutions to \eqref{eq: transfer matrix final solution} for $k_+=0$, and since there are $N$ solutions in \eqref{eq: k_+=0 solution} we have thus found all possible $k_+=0$ solutions.  We also note a general solution for any $k_+$ and $a$ Even is $E=-(1-\cos(\frac{a\pi}{N})\cos (k_+))$.  Out of all the solutions in the one spin-up sector, the largest non-zero solution is therefore $E=-(1-\cos(\frac{\pi}{N-1}))$ which provides the gap given in \eqref{eq: single particle gap}. 

\subsection{Numerics}
\begin{figure}
    \centering
    \includegraphics[width=1\textwidth]{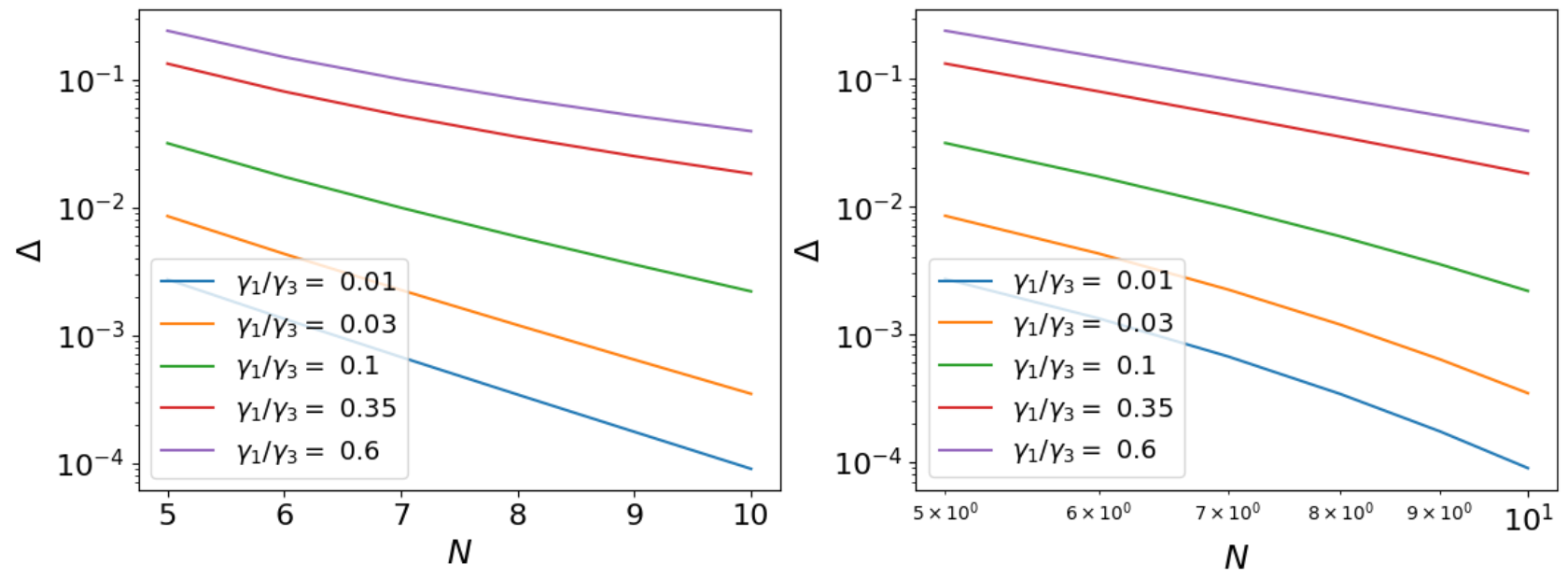}
    \caption{Decay of gap for increasing system size.  In comparing the log-linear plot (left) and log-log plot (right), a transition from exponential to power law decay appears as $\gamma_1/\gamma_3$ is increased.}
    \label{fig:pow vs exp}
\end{figure}

\begin{figure}
    \centering
    \includegraphics[width=0.5\textwidth]{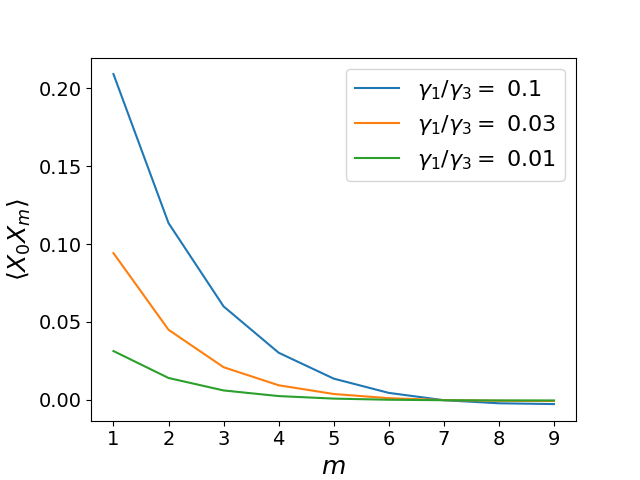}
    \caption{Decay of correlations for eigen-density matrix corresponding to the first non-zero eigenvalue.  }
    \label{fig: correlation decay}
\end{figure}

Below are small system size numerics supporting the NN results discussed in the main manuscript.  Namely, Fig. \ref{fig:pow vs exp} shows the transition from exponential gap to power law gap scaling in system size.  Fig. \ref{fig: correlation decay} shows correlations in the mixed phase.  Similar to the all-to-all case, the NN model exhibits correlations in the mixed phase which become stronger as $\gamma_3$ approaches the critical value.  However, in this case the correlations are short range, decaying exponentially with distance.  Numerics were performed on a system of 10 sites with open boundary conditions.  Correlations are for the eigendensity matrix corresponding to the mixed steady state.  As a technical point, the true steady state is a linear combination of the eigendensity matrices corresponding to the three 0 eigenvalues.  Since the eigenvalues are degenerate, one must take a general linear combination $\rho$ of the three steady eigendensity matrices and orthonormalize such that $\Tr{\rho} = 1$, $\Tr{\rho |W\rangle \langle W|} = 0$, $\Tr{\rho |0 \rangle \langle 0|} = 0$, and such that $\rho$ is positive semi-definite to find the mixed steady state.  

\begin{figure}
    \centering
    \includegraphics[width=0.99\linewidth]{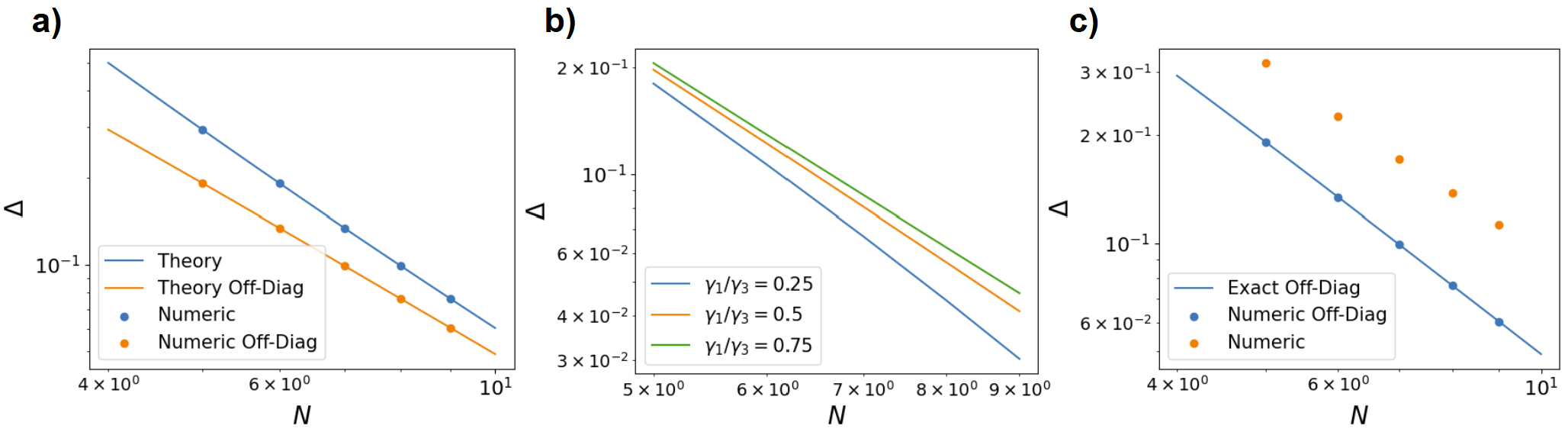}
    \caption{Gap and dynamical exponent in the ordered, critical, and disordered phases.  a) Gap in the W-state phase.  Exact analytical results for both the off-diagonal and physically relevant gap match that found by numerics.  b) Scaling of gap at critical point. c) Gap scaling deep in the mixed phase.  Numerical estimate of physically relevant gap plotted, and analytic off-diagonal result matches numerics.}
    \label{fig:Dynamical Exponent}
\end{figure}

We also find numerical estimates for the dynamical exponent deep in the W-state phase, the mixed phase, and at the critical point.  Results for dynamical exponents are supported numerically in two ways: first by exactly diagonalizing the Lindbladian for small system sizes, and second by evolving the magnetization density with time starting from the fully polarized spin-up state---as at long times, $\langle M(t) \rangle - \langle M \rangle_{\text{steady-state}} \sim e^{-\Delta t}$.  In both cases, periodic boundary conditions are used.  The results are plotted in figure \ref{fig:Dynamical Exponent}.  An important caveat is that the dynamics of the NN model does not generate any correlations between the different total $S_z$ sectors.  Thus, there will be eigenvalues corresponding to eigendensity-matrices with such correlations which play no role in the dynamics (unless initial states are chosen with these correlations).  Examples of such eigenvalues are given in section \ref{Sec: exact solution to off diag correlations with dark}.  In fact, the gap from \ref{Sec: exact solution to off diag correlations with dark}, $\Delta = (\gamma_1 + \gamma_3) \left[1 - \cos(\frac{\pi}{N}) \right]$, is exactly the gap found numerically for the W-state and mixed phases.  In figure \ref{fig:Dynamical Exponent} we label both the analytical solution and numerical evaluation of the eigenvalues corresponding to these eigendensity-matrices which are not block diagonal in the $S_z$ sectors as "Off-Diag."  The gap found via the dynamics of the system, i.e. extracted from the exponential decay of magnetization, does not match these eigenvalues.  However, when off-diagonal eigenvalues are excluded, the gap \textit{does} match the gap found from the magnetization decay.  In figure \ref{fig:Dynamical Exponent}, only the gap found from magnetization decay is plotted, as the gap found by exact diagonalization---after excluding the off-diagonal eigenvalues---matches within sufficient precision to be indistinguishable on the graph (all such gaps matched with $\geq 4$ significant figures).  Deep within the W-state preparation phase, this more physically relevant gap was found exactly in Sec. \ref{sec: single spin solution} to be $\Delta=-(1-\cos(\frac{\pi}{N-1}))$ and matches the numerics for small system sizes as shown in \ref{fig:Dynamical Exponent}a.  

From \ref{fig:Dynamical Exponent}a) it is possible to extract the dynamical exponent from the slope of the log-log plot.  The numerics give $z=1.96$ and $z=2.29$ for the off-diagonal and physically relevant gaps respectively.  The exact solution, however, scales as $z=2$ for large $N$ in both cases.  The difference between the numerical and exact solution highlights that at these system sizes only a very rough estimate of $z$ is achievable.  No exact solution was found in the mixed phase, but the relevant gap is shown in \ref{fig:Dynamical Exponent}c for numerically accessible system sizes.  Here we find $z=1.76$ for the physically relevant gap.  System sizes are too small to determine for certain if this $z$ is distinct from that found in the W-state preparing phase.    

Fig. \ref{fig:Dynamical Exponent}b shows the gap near the critical point.  In this case, the relevant gap closes and so is much closer to $0$ than the off-diagonal solutions.  For $\frac{\gamma_1}{\gamma_3}=0.5$ (which we take to be the approximate critical point), we find that $z = 2.66$.  It is hard to assess the accuracy of this critical exponent, as the small system size prevents a precise estimation of the critical point and---even if the critical point was known exactly---the numerical estimate of $z$ at these system sizes is imprecise (as seen comparing the exact and numerical values for $z$ in Fig. \ref{fig:Dynamical Exponent}a).  However, it is possible to bound $z$ by noting that $\gamma_3$ terms may only slow down preparation of the W-state.  This implies that the critical $z$ must be greater than or equal to the dynamical exponent for $\gamma_3=0$, i.e. $z \geq 2$.  This is sufficient to show that the $z$ is distinct from the DP 1D result, $z=1.5807$ and suggests the critical behavior is not of the DP type.        

\section{Coherent Terms}
It is possible to make the W-state the unique (absorbing) dark state through the addition of a pairwise, albeit non-local, Hamiltonian term.  One such example, introduced in \cite{Ticozzi2014DStatePrep}, is given by

\begin{gather}
    {\cal H} = \frac{1}{\sqrt{2}} \left(\sigma^x_1 \Pi_0 - \Pi_0 \sigma^x_N \right), \;\; \Pi_0 = \frac{1}{2} \left[ \sum_{i=2}^{N-1} \sigma^z_i - (N-4)I \right].
\end{gather}
It may be shown that the action of $\cal H$ on the vacuum and W-state are given by
\begin{gather}
    {\cal H}|W\rangle = 0, \;\; {\cal H}|\downarrow \downarrow\ldots \downarrow\rangle = \frac{1}{\sqrt{2}} \left( |\uparrow\downarrow\downarrow\ldots \downarrow\rangle - |\downarrow\ldots\downarrow\uparrow\rangle\right).
\end{gather}
This allows for evolution out of the vacuum while leaving the W-state dark.

A log-log plot of the gap for system sizes up to $N=10$, similar to Fig. \ref{fig:Dynamical Exponent}, yields $z\approx 2.77$ when $\gamma_3=0$ for the nearest-neighbor Lindblad evolution.  This value lower bounds $z$ for $\gamma_3>0$ and is thus consistent with a critical $z>2$.  A more rigorous analysis of the system size dependence of the gap when $\gamma_3=0$ as well as investigation into the nature of a possible transition when $\gamma_3>0$ is left for future work.     

\end{document}